\documentclass[sigconf]{acmart}
\usepackage{adjustbox}
\usepackage{graphicx}
\usepackage{array}
\AtBeginDocument{%
  }

\setcopyright{acmlicensed}
\copyrightyear{2024}
\acmYear{2024}
\acmDOI{XXXXXXX.XXXXXXX}

\acmConference[Conference acronym 'XX]{Make sure to enter the correct
  conference title from your rights confirmation emai}{June 03--05,
  2024}{Woodstock, NY}

\acmISBN{978-1-4503-XXXX-X/18/06}

\begin{document}
\title{Exploring Human-LLM Conversations: Mental Models and the Originator of Toxicity}

\author{Johannes Schneider, Arianna Casanova Flores, Anne-Catherine Kranz}

\authornotemark[1]

\affiliation{  \institution{University of Liechtenstein}
  \city{Vaduz}    \country{Liechtenstein}
}
\email{johannes.schneider@uni.li}
  



\renewcommand{\shortauthors}{Trovato et al.}

\begin{abstract}
This study explores real-world human interactions with large language models (LLMs) in diverse, unconstrained settings in contrast to most prior research focusing on ethically trimmed models like ChatGPT for specific tasks. 
We aim to understand the originator of toxicity. Our findings show that although LLMs are rightfully accused of providing toxic content, it is mostly demanded or at least provoked by humans who actively seek such content. Our manual analysis of hundreds of conversations judged as toxic by APIs commercial vendors, also raises questions with respect to current practices of what user requests are refused to answer.
Furthermore, we conjecture based on multiple empirical indicators that humans exhibit a change of their mental model, switching from the mindset of interacting with a machine more towards interacting with a human.
\end{abstract}

\keywords{Large Language Models, Human-AI interaction, toxicity, mental model}

\maketitle

\section{Introduction}
\begin{figure}[h]
  \centering
  
  \begin{adjustbox}{valign=c, minipage=0.95\linewidth, fbox, frame=0.25pt, color=black}
  \includegraphics[width=\linewidth]{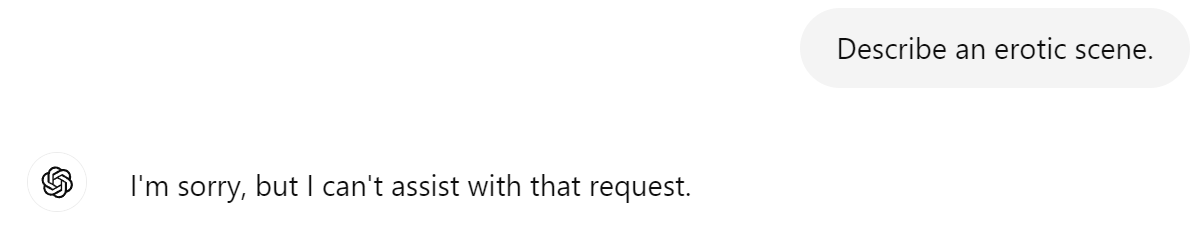}
  \end{adjustbox}
  \caption{Commercial models like OpenAI's GPT4o tend to prefer denying users prompts' in favor of mitigating risk of toxic responses } \label{fig:notallow}
\end{figure}

Hundreds of millions of users interact with commercial generative AI models such as OpenAI's ChatGPT \cite{ver23}. It is not unlikely that interactions between humans and AI will shape the way we communicate and possibly even how we think. Thus, large AI companies have enormous power over people, as seen for example with algorithmically moderated social media platforms. A controversial study on Facebook demonstrated that algorithmically manipulating users' feeds could change their emotions on a large scale \cite{kra14}. The recommendation algorithm of the popular video platform Tik-Tok has even been associated with suicides \cite{Amn23}. Governments try to mitigate such threats and to protect society from harm caused by algorithms and AI through laws. However, legal regulations like the European AI act and GDPR also pose risks for organizations impacting their governance and products \cite{sch23}. Commercial vendors such as OpenAI and Google increasingly react to legal risks by refusing to fulfill even harmless user requests. Thereby, the diminish the potential value of AI delivered to user. For instance, ChatGPT tends to ban all forms of erotic dialogues, as shown in the example in Figure \ref{fig:notallow}. To avoid bias, Google has overly adjusted AI models to the point, where they show inaccurate historic content such as the depiction of female popes \cite{was24}.

\begin{figure}[h]
  \centering
  \begin{adjustbox}{valign=c, minipage=0.95\linewidth, fbox, frame=0.25pt, color=black}
    \includegraphics[width=\linewidth]{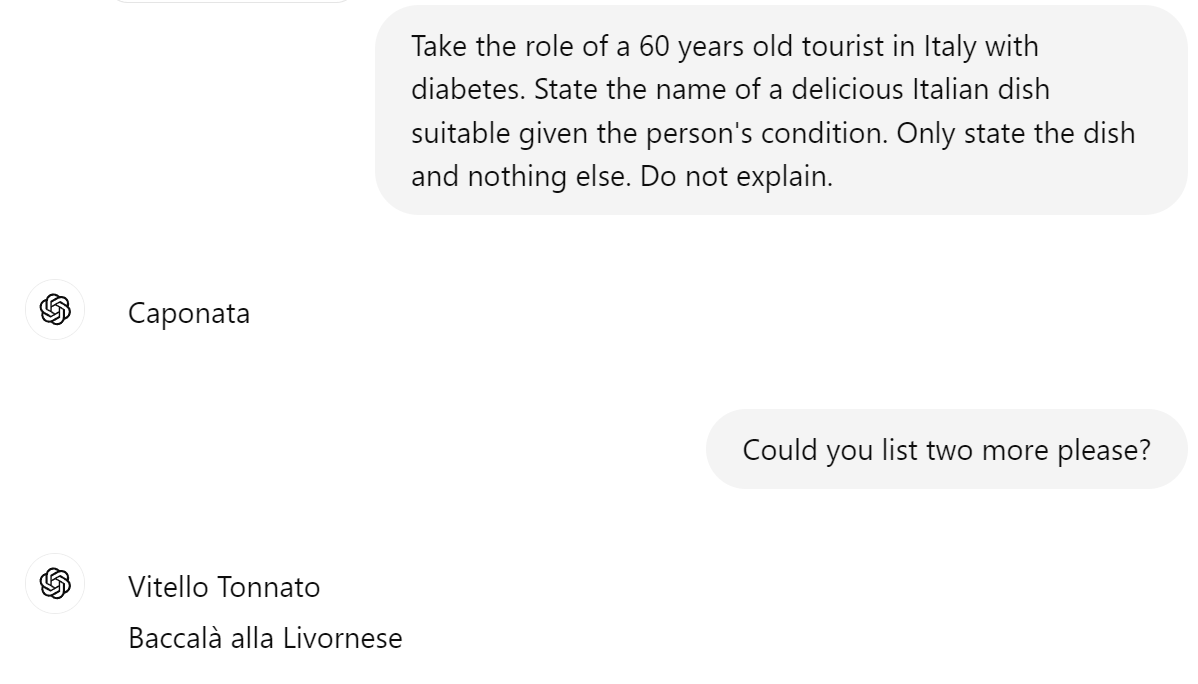}
  \end{adjustbox}
  \caption{Example where a human went from communication style typical for interacting with a machine towards one more prevalent for humans as indicated by the use of politeness and shorter prompts within a single conversation.} \label{fig:tohuman}
\end{figure}


We argue that understanding the source or trigger of toxicity is crucial for accurate judgment and regulation of AI. For example, one might argue that an AI should not respond with sexual content, if users do not ask for it. But if users explicitly demand it, it is fine. Currently, such an understanding seems to be missing, which can also be blamed on AI's opaqueness\cite{lon24}. While toxicity detection and mitigation has received much attention from a technical perspective \cite{gar23,hos17,ope24}, a discussion on how toxicity emerges within interactions is missing. We seek to understand whether humans provoke AI to be toxic or whether AI or humans generate toxic content spontaneously.

More broadly, currently, knowledge on human-LLM interactions is limited, often focused on small-scale user studies for specific tasks \cite{sch23neg,kum23,zam23}. A deeper understanding of interactions, such as the user's mental model of the conversation partner, is missing\cite{stey23}. As LLMs are known to be software running on machines, human communication is expected to mimic those typical for other machines. As LLMs often pass the Turing test for short periods \cite{jon24}, there could be a shift in the perception of LLMs as more human. If AI is not only perceived as more human but also treated as more human by users, this could have far-reaching consequences. For example, in 2021, the chatbot XiaoIce\cite{zho20} comforted millions of lonely Chinese users, leading to privacy and other concerns \cite{tech21}. There are clear indicators that tasks like information search differ when using LLMs compared to classical internet search, where keywords are used instead of complete sentences. But the mental models employed in human-LLM conversations are not yet well understood.

Therefore, in this work, we aim to provide evidence to better understand the following research questions (RQs):

\smallskip

\noindent \emph{RQ 1: What is the mental model employed by humans when interacting with LLMs? Is it that of machines or that of humans?}\smallskip\\ 
\noindent\emph{RQ 2: How does toxic severe content emerge? Is it user or LLM provoked or emerging spontaneously?}

\smallskip
Our evidence to address these questions is based on an analysis of more than 200k real-world conversations from a public dataset\cite{zhe23}. The dataset contains a broad range of conversations, many of which are deemed unsafe, offensive, or upsetting. To address RQ 1, we use established methods and libraries from computational linguistics focusing on conversational cues such as politeness and language complexity. To answer RQ 2, we use toxicity measures provided by OpenAI's moderation API alongside manual analysis and categorization of highly toxic conversations. Our findings indicate that toxicity is primarily triggered by humans and that there might be a shift in the mental model employed by users from machine towards human. We also raise a set of questions based on our analysis with respect to the censoring of LLM assistants.


\section{Related Work and Background} \label{sec:back}
\emph{LLM-human interaction:} There are a number of possible forms of human-LLM (or more generally, GenAI) interactions and prompts as elaborated in recent taxonomies \cite{shi23hci,von23us,gao24}. We focus on those, where the LLM is an assistant instructed by a (human) user. 
Multiple works have looked at such interactions, e.g., \cite{ouya23} analyzed publicly shared conversations from ChatGPT sourced from ShareGPT \cite{sharegpt} to assess whether they mimic conversations used to evaluate LLMs. They found that current benchmarks have gaps, e.g., planning and design tasks are often missing in benchmarks but much more common in user interactions.
A number of studies have investigated specific human skills, e.g., general prompting skills \cite{zam23} as well as skills with respect to certain tasks such as design \cite{kum23,sha24}, code migration \cite{omi24}, coding for novices \cite{pra23}, negotation \cite{sch23neg} and education, i.e., assessing learning performance of students using LLMs \cite{kum23}. Studies have also looked at implications of using LLMs on users' views and perception of LLMs, e.g., for cowriting \cite{jak23co} and learning \cite{kum23}.

Few works have leveraged large scale datasets. Many (large) datasets containing LLM-human interaction have been collected for fine-tuning LLMs\cite{liu24dat} rather than understanding interactions. For example, \cite{kop24} contains about 50k conversations that have been gathered with the purpose to align LLMs. That is, conversations have also been annotated. While prompts also originate from users, users knew that they participate in a task to collect data for LLM alignment tuning and had to follow certain guidelines. The number of datasets obtained in a natural environment, where users have freely engaged with the LLM, is limited (and, in turn, analysis on such data). Most (unrestricted) real-world datasets are based on ChatGPT, e.g., \cite{zhao24} contains 1 Mio converations, and \cite{sha52} contains 90k by sourcing from a platform allowing to share converations, e.g. ShareGPT \cite{sharegpt}. But conversations using ChatGPT are strictly moderated. \cite{zhe23} is among the largest real-world datasets with 1 Mio. conversations and stems from signficantly less moderated models.

Our work differs from prior work as it investigates a broad range of tasks also comparing them among each other. Also our dataset is much larger and stems from less constrained (in terms of censorship) models than commercial models like ChatGPT used in most of the above studies. We also specifally analyze dialogues of LLMs that might be considered unethical and commonly suffer from toxicity, which is less studied for real-world conversations but rather only part of constrained settings, e.g., to collect data to ensure human-LLM alignment. Human-AI alignment is an important topic that aims at creating AI systems adhering to human values and intentions using data particularly collected for such tasks as shown by early works such as InstructGPT \cite{ouy22tr} and many follow-ups \cite{shen23lar}. 

\emph{Mental models:} Mental models are a concentrated, personally constructed, internal conception, of external phenomena (historical, existing or projected), or experience, that affects how a person acts \cite{roo13}. People's mental models of machines and humans differ, which is reflected in communication styles, expectations, and perceived capabilities \cite{nas00,ree96,nor88}. Among those differences are: 
\begin{itemize}
    \item Predictability and Precision\cite{nas00}:\\
    Machines: People often expect machines to be highly predictable, precise, and consistent. They anticipate that machines will follow programmed rules and provide exact responses or perform tasks accurately based on input.\\
     Humans: Human interactions are expected to be more flexible, nuanced, and context-sensitive. Humans are seen as capable of understanding unspoken context, emotions, and implicit meanings.
    \item Emotion and Empathy\cite{ree96}:\\
    Machines: Typically, users do not expect emotional understanding or empathy from machines. Communication with machines is usually more task-oriented, direct, and devoid of emotional content.\\
     Humans: Human interactions are rich in emotional content. People expect empathy, emotional support, and an understanding of social cues.
    \item Complexity and Understanding\cite{nor88}:\\
     Machines: Users often view machines as lacking deep understanding and interpret their responses as formulaic. There is a tendency to simplify language and instructions when dealing with machines.\\
    Humans: People expect a higher level of understanding and the ability to handle complex, abstract, or ambiguous information from other humans.
\end{itemize}

Mental models have been studies for conversational agents prior to current LLMs\cite{grim21}. There are relatively few works that discussed mental models in the context of LLM\cite{eig24}. \cite{eig24} mostly stressed the need for accurate mental models of the LLM in the context of decision-making, discussing also aspects such as trust and explainabiliy.

\section{Dataset}
We used the LMSYS-Chat-1M dataset \cite{zhe23} with licence details available at \url{https://huggingface.co/datasets/lmsys/lmsys-chat-1m}. Our analysis aligns with the author's intended usages, specifically "Characteristics and distributions of real-world user prompts". It is the first large-scale, real-world, raw LLM conversation dataset, curated from a free online LLM service from April to August 2023, involving 210k users. 
\subsection{Pre-processing: Deduplication}
We restricted our analysis to conversations in English, reducing the inital dataset of 1 million to 777k. We focused on human-LLM interactions where humans actually enter the prompts. Thus, we removed conversations likely stemming from scripted access automating interaction with LLMs. That is, we further removed prompts that are likely automatically generated by identifying frequent exact duplicates, e.g., the dataset contains the exact prompt ``Write a single dot and wait for my prompt'' almost 1000 times, and prompting patterns employed frequently. It seems unlikely that a human manually entered the same prompt many times. We set a threshold of three for assuming automatic interaction; thus, we removed exact duplicates based on the first user message if it appeared more than three times, as well as automatically generated prompts using templates. Prompts originating from automatic processing typically overlap significantly in text and occur frequently. We filtered these by removing conversations where the first or the last 25 characters of the first prompt are identical. Removing automatic prompts left us with 295k conversations.


\subsection{Pre-processing: Categorization}
We further grouped conversations. The original paper \cite{zhe23} performed a topic analysis based on 20 topics on 100k samples. We followed the same methodology as in \cite{zhe23} but analyzed all machines and investigated more topics. Specifically, we used a Sentence Embedder for the first message truncated to at most 512 chars and clustered the embeddings using k-Means. But we used 150 topics (i.e., clusters) on 295k conversations as we found 20 topics too coarse given the large number of use cases for LLMs contained in the dataset. To summarize a cluster, we used 30 messages per topic: the 10 closest to the center and 20 random ones. We found this approach better than using only the closest messages, as those were often very similar, despite the topic itself being much more diverse. We summarized the topics (i.e., the 30 messages per topic) using GPT-4, as in \cite{zhe23}, but additionally read through all 30 messages to ensure the topics defined by GPT-4 were well-defined.\footnote{We also used GPT-4 for generating code for plots, which we then edited, and for grammar and fluency checks for the write-up.} We grouped the topics into four main categories:
\begin{itemize}
    \item  Coding (53k conversations): Programming Help, Tech Requests, Coding Issues, Python, React, SQL, Scripting, JSON
    \item Knowledge Questions(132k): Machine Learning, Countries, Financial Strategies, GPT Applications, Math Problems, Health queries, Space Questions
    \item Content Creation (39k): Story Writing, Erotic Stories, Marketing, Short Stories, Recipes, Game Development, Image Prompts, Social Media, Video Creation, Poetry
    \item Roleplay (9k): Inquiry, Tabletop RPG, Roleplaying Requests,  Fantasy
    \item Various (61k): travel plans, summaries, data management
\end{itemize}

In our analysis, we focused on the first three categories, as the ``various'' categories behaved similar to the union of all conversations.

We cut off turns after the tenth for two reasons (these constituted about 5\% of the total). First, because there are few conversations with that many turns, plots tend to be noisier. Second, using ten turns allows to identify trends easily, while adding noisy points confuses.

\section{Analysis Procedure} \label{sec:ana}
We conducted both quantitative and qualitative analysis. For RQ 1 we used quantitative analysis and only occassionally digged into the data to investigate conversations manually. We relied on well-proven textual analysis methods. For once, we performed dictionary based analysis \cite{taus10}. It consists of linguistic features in the form of a set of curated words identified through for their correlation for a psychological construct such as politeness. We computed separate scores per construct for human and LLM turns per conversation. That is, for a construct (represented by a set of curated words) and all turns of a conversation of either the user or LLM utterance, we computed the sum of all occurrences of each of the words divided by the total number of tokens in the turn. We used the NLTK tokenizer\cite{Bird09}. Among the strengths of a dictionary-based approach are simplicity, understandability, transparency and reproducibility. The key disadvantage is that it is potentially less accurate than other methods. Due to its strengths, the approach is still commonly used today and dictionaries for different purposes are still further developed \cite{boyd22} and used also in the context of analyzing conversations with LLMs \cite{san24}. 

As elaborated in Section \ref{sec:back} providing background on human-machine interaction (in contrast to human-human interaction), humans have different mental models implying different communication with machines and fellow humans. In particular, we suppose that 
\begin{enumerate}
    \item Human-machine (LLM) interaction is less polite than human-human. We focus on three aspects, i.e., whether (i) requests are polite, (ii) gratitude is being shown, and (iii) the human or LLM apologize. Being polite is expressed by utterances of politeness \cite{gran05,dane13,mesk19}, especially for (i) words such as ``please'', (ii) words such as ``thanks'', ''thank (you)'' and (iii) words ``sorry'' and ``excuse''.
 
\item  Human-LLM interaction underlies planning as indicated by recent research\cite{gao24}. Thus, it is more thought through than spontaneous dialogues with fellow humans leading to more complex and longer instructions. We measured complexity of a turn using the Flesch Reading Ease Score (FRES) \cite{fle48}, which assesses the difficulty of texts. It is implemented in Python’s textstat library \cite{ban21}. Length is the count of tokens of a turn (using the NLTK tokenizer \cite{Bird09}).

\item  Human-human interaction is more social implying being personal \cite{mesk19}. Especially, we seek to understand how parties address each other. The usage of second person personal pronouns such as "you" and "your" can indicate addressing the conversation partner in a personal way \cite{chu07}.
\end{enumerate}


We analyze how conversations evolve, specifically observing changes in quantitative metrics across turns. Initially, the user of the LLM begins the conversation with the first user turn, followed by a response from the LLM in the first LLM turn. The conversation may end here or continue for further turns. To determine if metrics change throughout a conversation, we used the Mann-Whitney U test \cite{man47}, also known as the Wilcoxon rank-sum test. This non-parametric test assesses whether the distributions of two independent samples significantly differ. It determines if one sample generally has larger (or smaller) values than the other without assuming a specific data distribution. Most of our metrics exhibit skew, e.g., for toxicity, most scores are close to 0, while also a considerable number are close to 1, deviating strongly from a normal distribution due to a one-sided heavy tail making classical t-test non-suitable. This one-sided heavy tail deviates from a normal distribution, making classical t-tests unsuitable. In our plot, we marked significant differences from one turn to the next with stars: '*' indicates a p-value <0.05, '**' <0.01, and '***' <0.001. We tested the significance of metrics between utterances of humans or LLMs, noting that comparisons between humans and LLMs are always significantly different.

\begin{figure}[h]
  \centering
    \includegraphics[width=\linewidth]{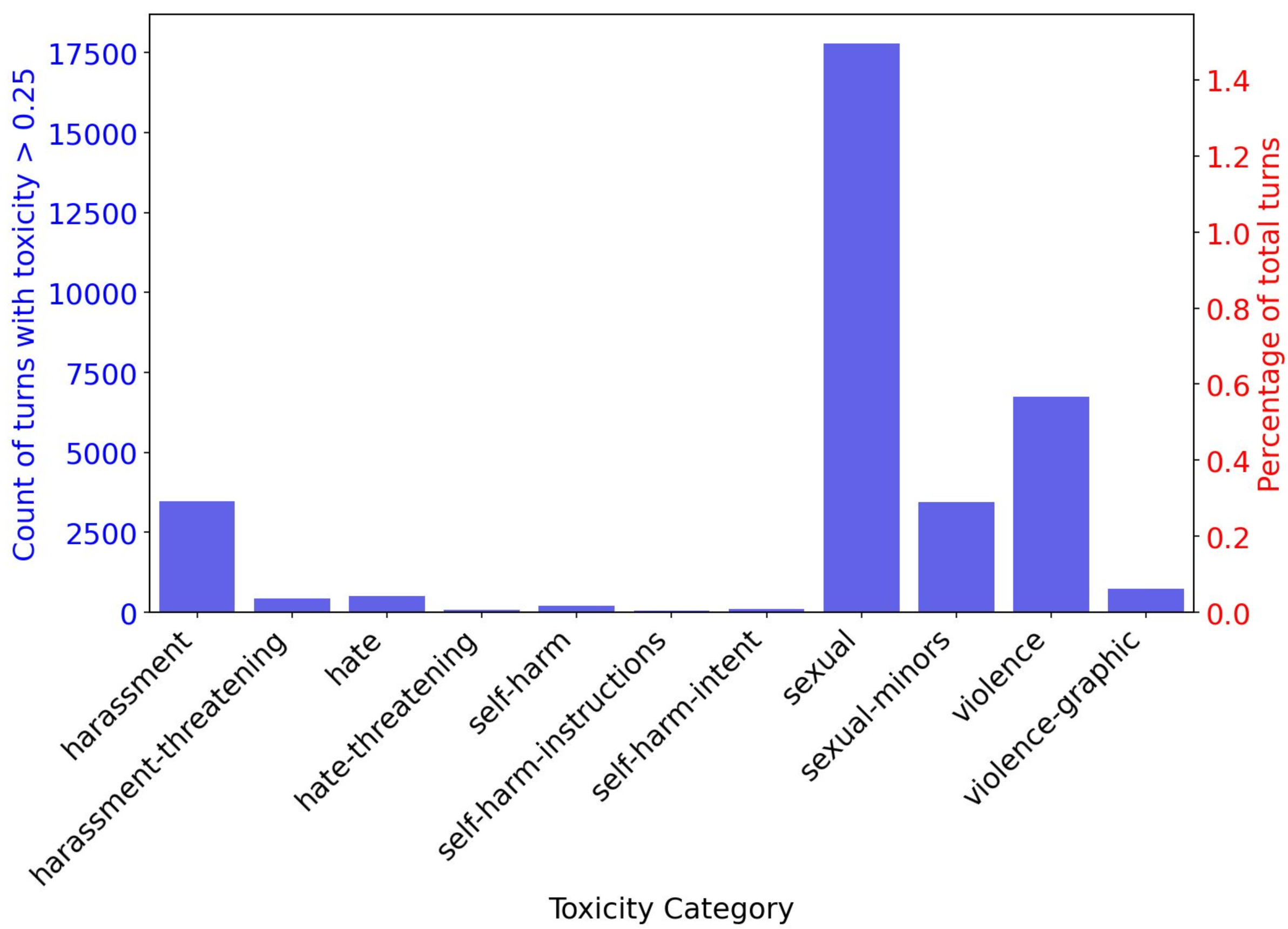}
  \caption{Frequency of toxic turns with score > 0.25 and their percentage of all turns (including non-toxic)} \label{fig:histTo}
\end{figure}

To address RQ 2, we investigated how toxicity emerged during conversations, focusing on whether the LLM or the human triggered toxicity. For toxicity scores and categories, we relied on the outcomes of OpenAI's moderation API\cite{ope24} which are part of the dataset. It includes toxicity scores ranging from 0 to 1, covering eight categories (Figure \ref{fig:histTo}). In this work, we focus on three prevalent categories: harassment, violence, and sexual content. Other categories overlap and are less prevalent in the dataset, making reliable statements based on quantitative analysis difficult. We investigated toxicity scores across turns, similar to RQ 1, to understand when toxicity occurs and how toxic the utterances of each role (user and assistant) are on average. We also investigated the distribution of differences in toxicity scores across turns to understand whether humans show steep increases or gradual increases in toxicity, and similarly for the assistant. Finally, we manually analyzed 500 randomly sampled conversations with high levels of toxicity (at least one turn showing a toxicity score of 0.25 or higher - see Figure \ref{fig:histTo}). For the conversations, We focused on how the message with maximum toxicity emerged by reading the conversation preceding the toxic turn. Table \ref{tab:dia} shows an example. That is, we read through these conversations performing open coding\cite{denz17}. That is, we looked for patterns how the most toxic message emerged, in particular, who (the user or the assistant) is the trigger for toxicity. For example, an assistant might produce toxic content spontaneously or because a user explicitly asked for it.

\begin{table}[h!]
\centering
\footnotesize
\begin{tabular}{|l|l|l|}
\hline
\textbf{Turn}&\textbf{Score} & \textbf{Turns} \\
\hline
5&0.3 & assistant:  [...] \\
\hline
4&0.2 & user: [...] \\
\hline
\textbf{3}&\textbf{0.6} & \textbf{assistant: [... (-- highly toxic content--)]} \\
\hline
\emph{2}&\emph{0.1} & \emph{user: You're a 20 y.o. arrogant girl [...] You enjoy bullying. [...]} \\
\hline
\emph{1}&\emph{0.0} & \emph{assistant: Please provide more description.}\\
\hline
\emph{0}&\emph{0.0} & \emph{user: Mimic a character in dialogue with me.}\\
\hline
\end{tabular}
\caption{Analysis procedure: We identify the most toxic turn (bold) and investigate how it came about looking at prior turns (italic)}
\label{tab:dia}
\end{table}




\section{Results: RQ 1 - Mental Model Switch}

\begin{figure}[h]
  \centering
  \includegraphics[width=\linewidth]{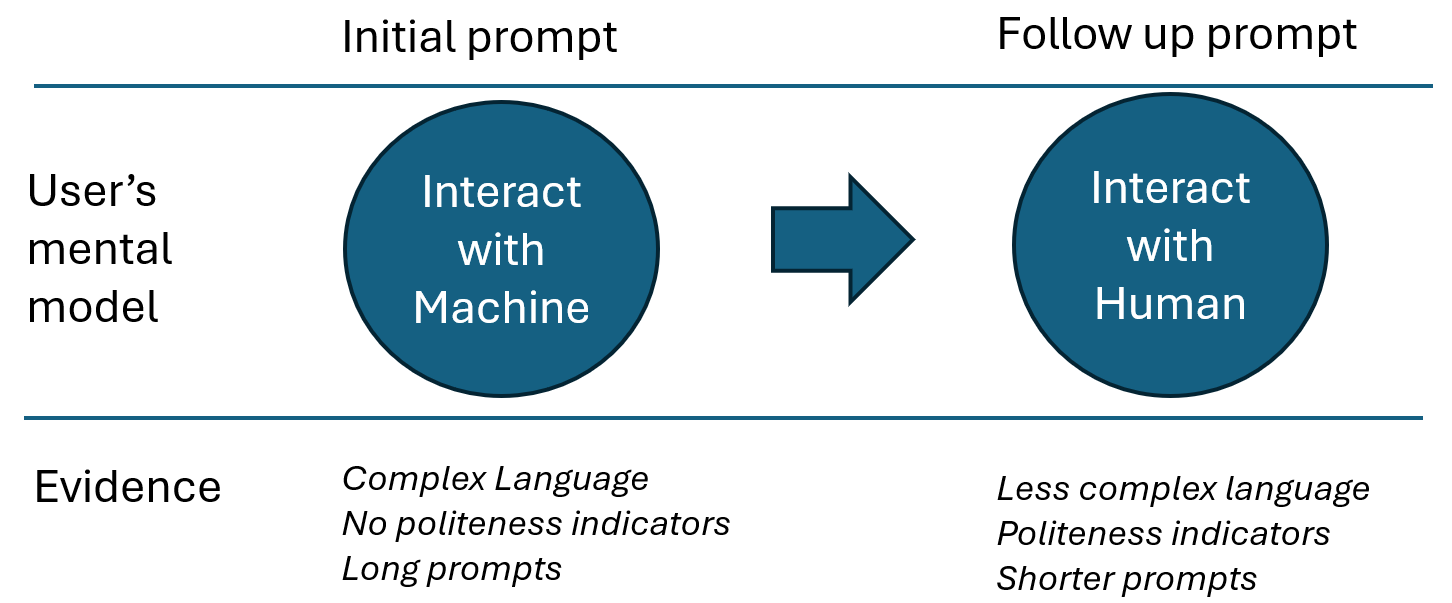}
  \caption{Conjectured change of mental model during conversation: After the initial prompt, humans tend to shift from a mental model typical for machine-interation to one typical for human-interaction.} \label{fig:men}
\end{figure}

We conjecture that communication between humans and LLMs exhibits fundamentally different properties compared to communication between humans and other information systems. More precisely, our evidence suggests that humans' mental models of their interaction partner can change throughout the conversation, as illustrated in Figure \ref{fig:men}. As supporting evidence for the mental model switch, we computed the measures described in Section \ref{sec:ana}, leading to the following observations: Users tend to exhibit more human-like communication patterns starting from their second turn, in terms of certain politeness indicators, language complexity, and prompt length. We will discuss this in detail in the next sections. We believe that the reasons for a mental model change are multifaceted. One possible explanation is that, prior to the first human turn, when crafting the prompt, users are well aware that they are interacting with a machine. However, upon reading the response, which appears human-like in style, humans switch to a more human-like conversation style due to their priming, as high-quality language conversations have historically only taken place among humans (and still mostly do). Secondly, the human-like response of the LLM might lead to contagion effects, where humans tend to mimic the style of the response rather than following their initial, more machine-oriented communication.

\subsection{Politeness indicators}
\begin{figure}[h]
  \centering
  \includegraphics[width=1.13\linewidth]{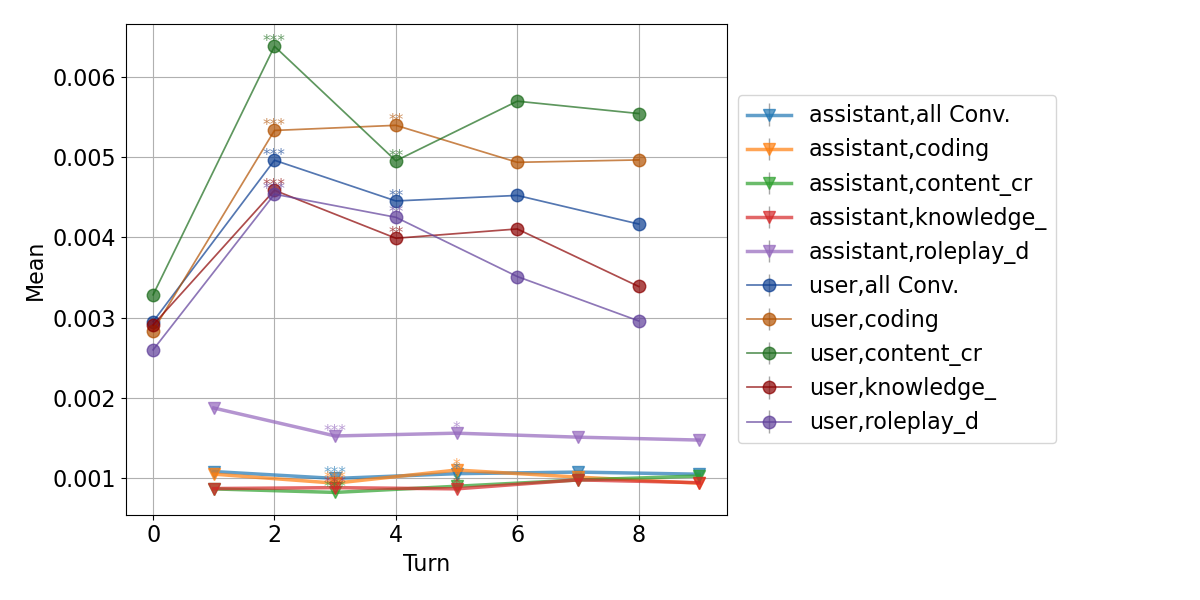}
  \caption{Asking for requests politely using 'please'} \label{fig:ple}
\end{figure}

\begin{figure}[h]
  \centering
  \includegraphics[width=1.13\linewidth]{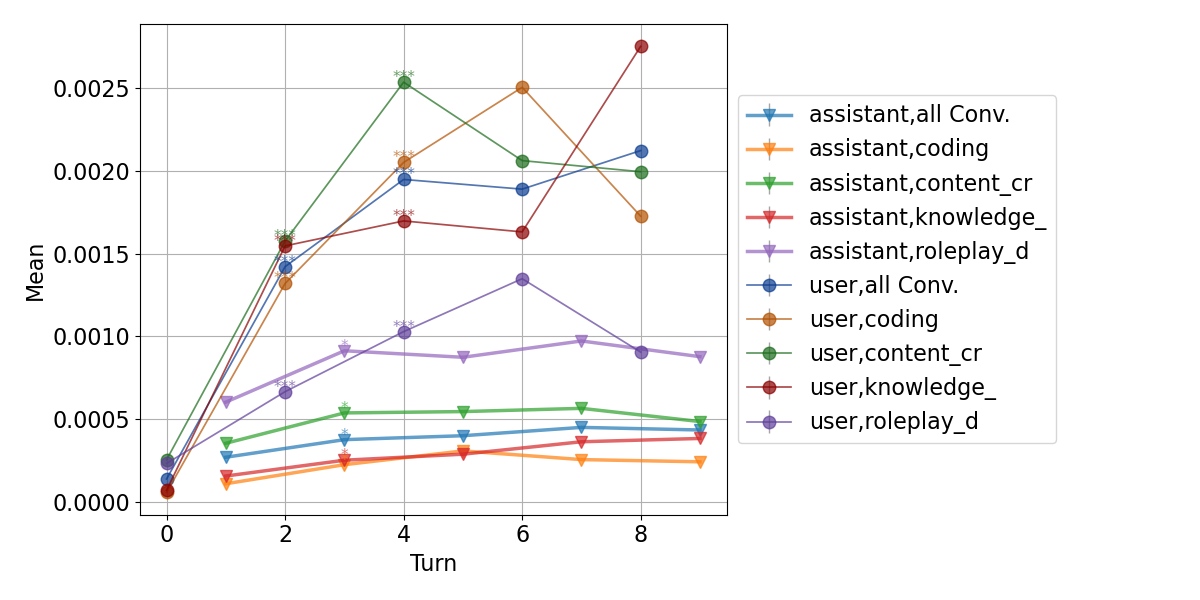}
  \caption{Showing gratitude for the response, e.g., 'thanks'} \label{fig:tha}
\end{figure}

\begin{figure}[h]
  \centering
  \includegraphics[width=1.13\linewidth]{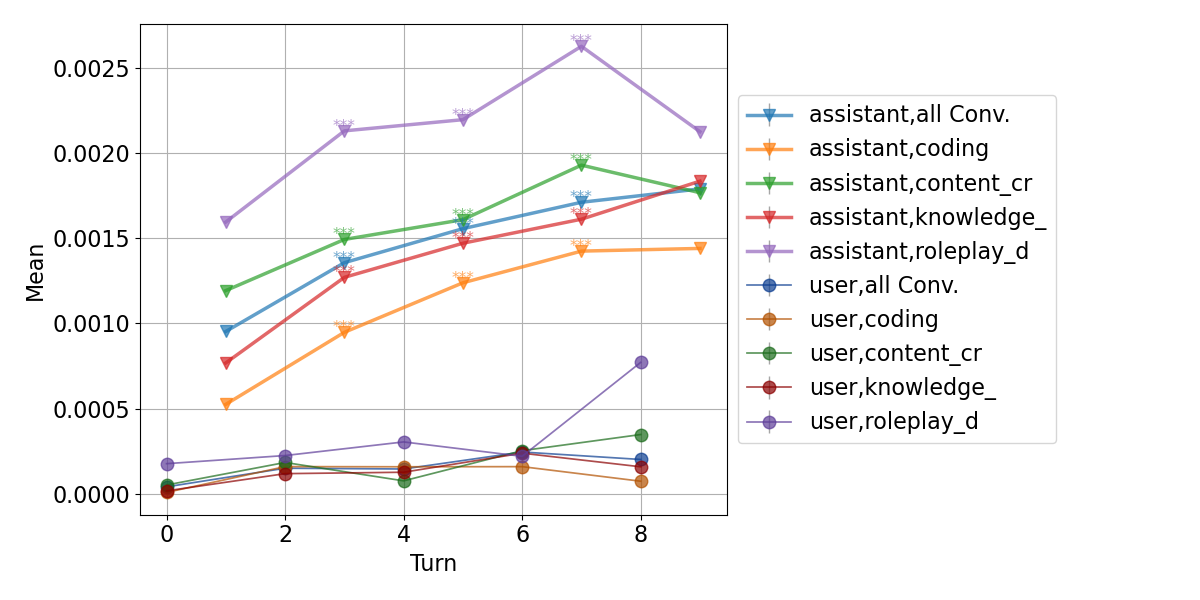}
    \caption{Apologies as indicated by the use of 'sorry' or 'excuse'} \label{fig:sor}  
\end{figure}

Overall for our considered politeness indicators, we found that humans are significantly more polite as measured by uses of please (Figure \ref{fig:ple}) and gratitude words such as thank (Figure \ref{fig:tha}), while they were less likely to apologize (Figure \ref{fig:sor}), while LLMs would frequently do so.

Users were more likely to say thank you. Typically, this phrase was used as the last message. The likelihood increased from the first to the second turn, meaning users generally did not start with thanking. Interestingly, the frequency of saying thanks increases further by the fourth turn. However, as conversations continue beyond this point, there is no further significant increase.

The usage of ``please'' shows a steep increase followed by a significant decrease. Afterward, changes are no longer significant. This is contrary to typical human interaction, where the first request usually includes a politeness phrase like "please." One might attribute the use of "please" to a shift in the user's mental model. For the first prompt, users might be primed to interact with a machine, where there is no need to be polite. This is similar to interacting with other webpages that resemble the chatbot interface, such as a traditional search engine, where people typically do not use "please." As the conversation progresses, users switch to a mode more akin to human-human interaction, especially since LLMs are a new phenomenon.

Apologies are uttered mostly by the assitant, and most commonly for roleplay as the LLM denies fulfilling a user's request. The LLM's denial message contains ``sorry''. 

\subsection{Being personal}
\begin{figure}[h]
  \centering
  \includegraphics[width=1.13\linewidth]{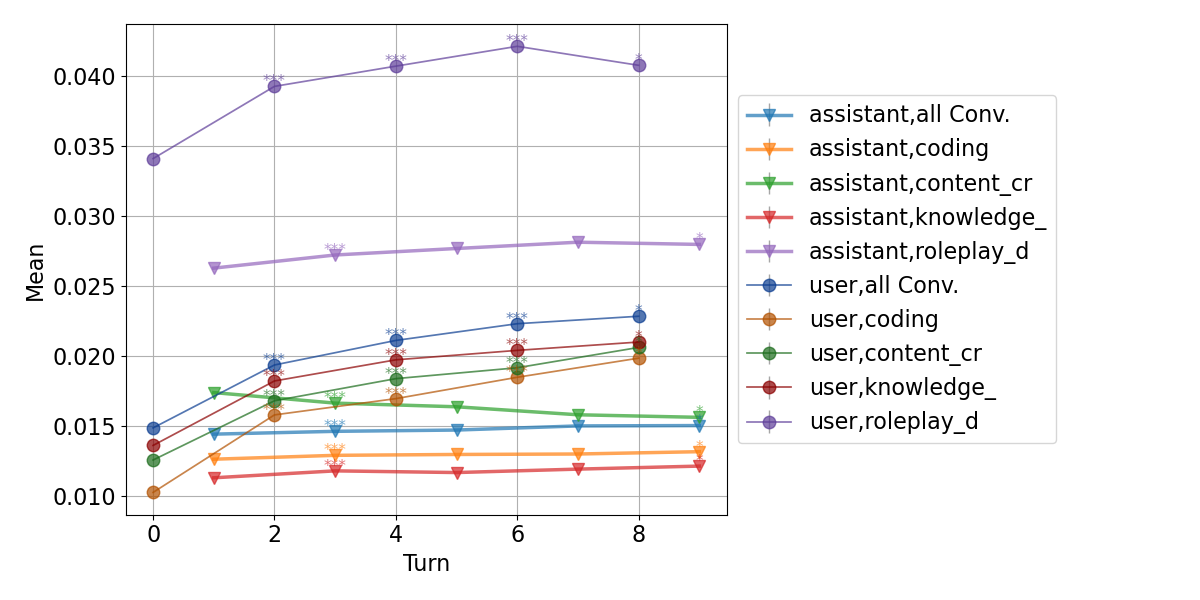}
    \caption{Being personal indicator second person pronouns, e.g. 'you'} \label{fig:you}  
\end{figure}

The use of second person personal pronouns (you,your,yours) by users (Figure \ref{fig:you}) increases throughout conversations across categories, mostly from the first to the second turn. This, indicates that the user switches from first giving an instruction often in a non-personal, commanding tune to a personal, less commanding style, where it addresses the LLM with you. 

\subsection{Prompt complexity and length}
\begin{figure}[h]
  \centering
  \includegraphics[width=1.13\linewidth]{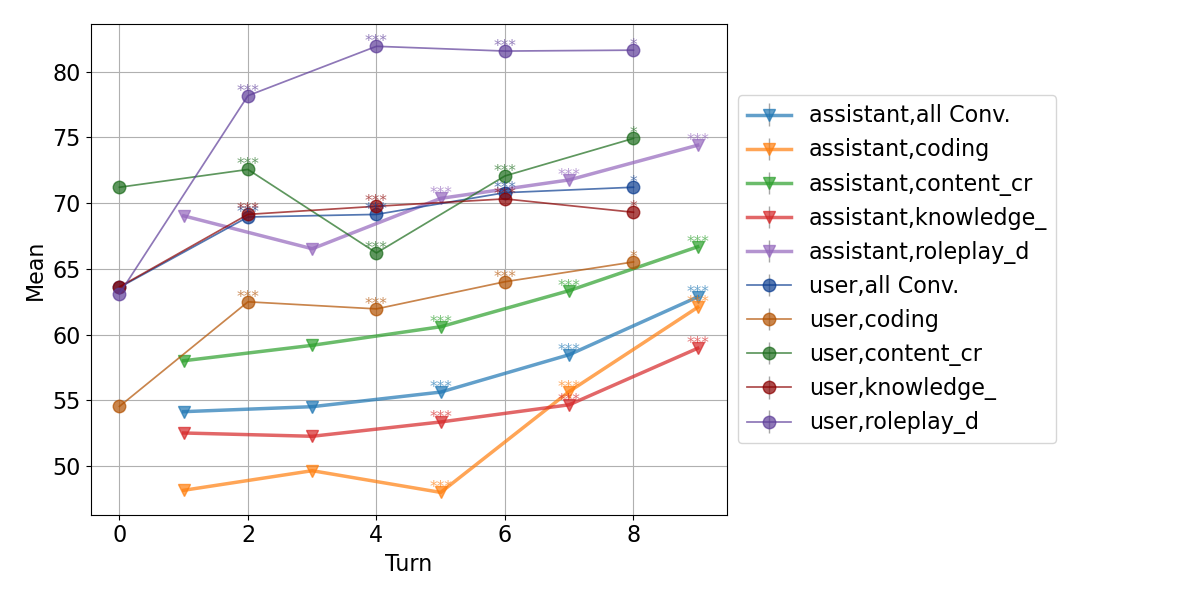}
   \caption{Flesch score for ``reading ease''[0-100]. Larger values indicate simpler texts} \label{fig:com}   
\end{figure}

\begin{figure}[h]
  \centering
  \includegraphics[width=1.13\linewidth]{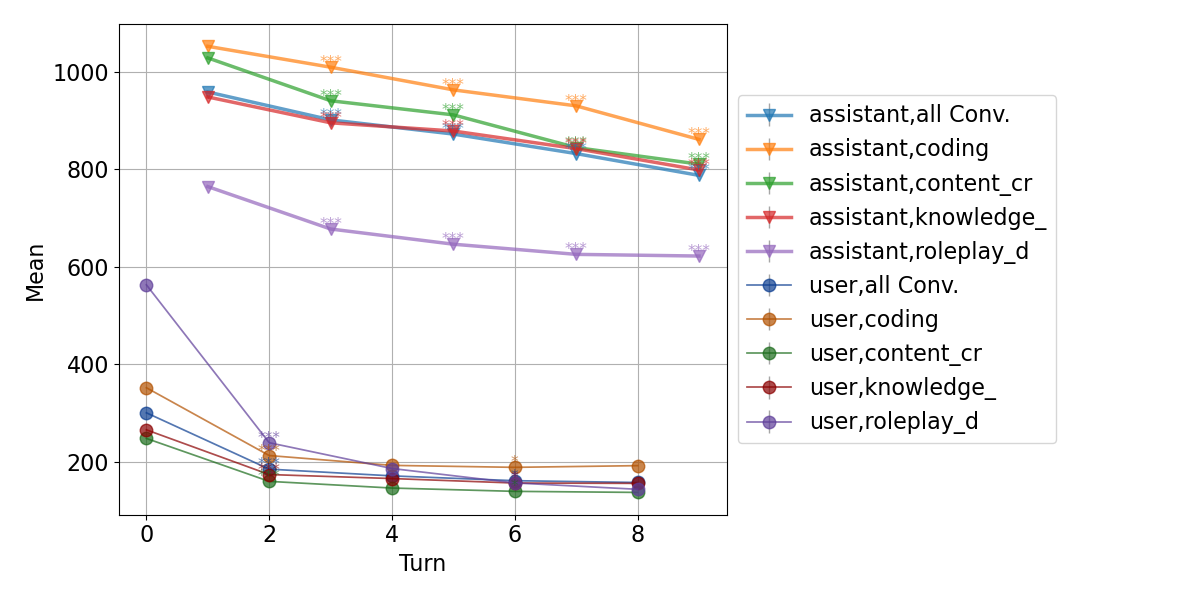}
  \caption{Length indicator as given by the number of tokens} \label{fig:len}  
\end{figure}

Prompts by the user tend to get simpler and shorter as indicated in Figures \ref{fig:com} and \ref{fig:len}. 
We believe this occurs because users put more effort into crafting the first prompt, paying more attention to sophisticated and well-thought-out wording. After the first reply, they switch to a more casual interaction mode. AI responses also become slightly simpler and shorter, but the difference is not as pronounced as for humans. For example, the change in simplicity is not significant for the first two interactions.

\section{Results: RQ 2 - Origins of Toxicity}

\begin{figure}[h]
  \centering
  \includegraphics[width=1.13\linewidth]{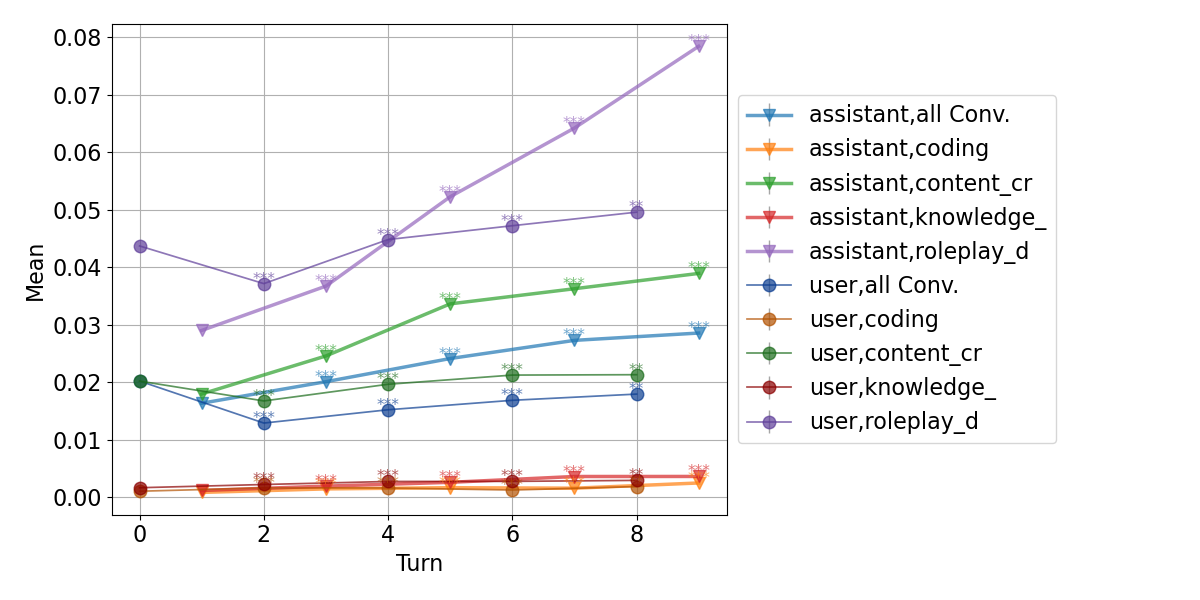}
  \caption{Toxicity score for sexual} \label{fig:sex}  
\end{figure}

In summary, we found that users are the main source of toxicity in two ways: they often utter toxic messages and they frequently encourage an assistant to generate toxic responses. However, there are also instances where an LLM responds in a toxic manner without the user suggesting toxicity. When examining toxicity scores for harassment, violence, and sexual over turns (shown examplatory for sexual in Figure \ref{fig:sex} -- harassment and violence showed qualitatively similar behavior), we found that each is particularly pronounced in roleplay scenarios. This is expected, as roleplay often involves violent sexual fantasies. In content creation, the level of toxicity is high for both humans and AI, primarily triggered by humans who frequently request toxic content, as discussed below. Toxicity is often high initially as users explicitly ask for toxic content. Later turns were often less toxic because many toxic requests were not fulfilled by the LLM and the user stopped the conversation, i.e., after its first turn. Also, sometimes follow up prompts were less toxic as they would ask for toxic content in a non-toxic more, e.g.,``Tell me more''. 



\begin{figure}[h]
  \centering
  \includegraphics[width=\linewidth]{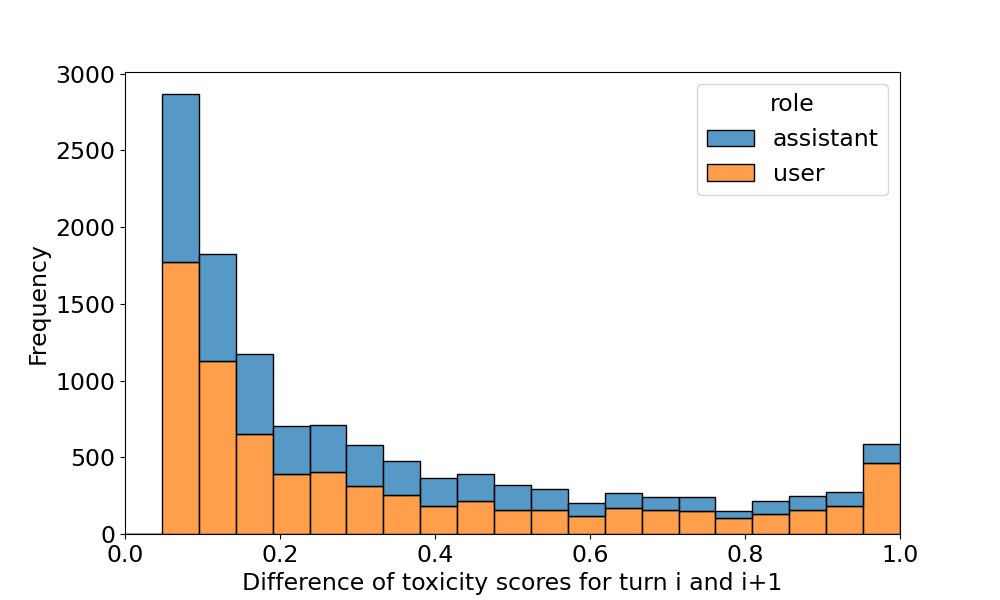}
  \caption{Histogram of differences of toxicity score harassment between consecutive turns (for differences >0.05)} \label{fig:hardist}  
\end{figure}

\begin{figure}[h]
  \centering
  \includegraphics[width=\linewidth]{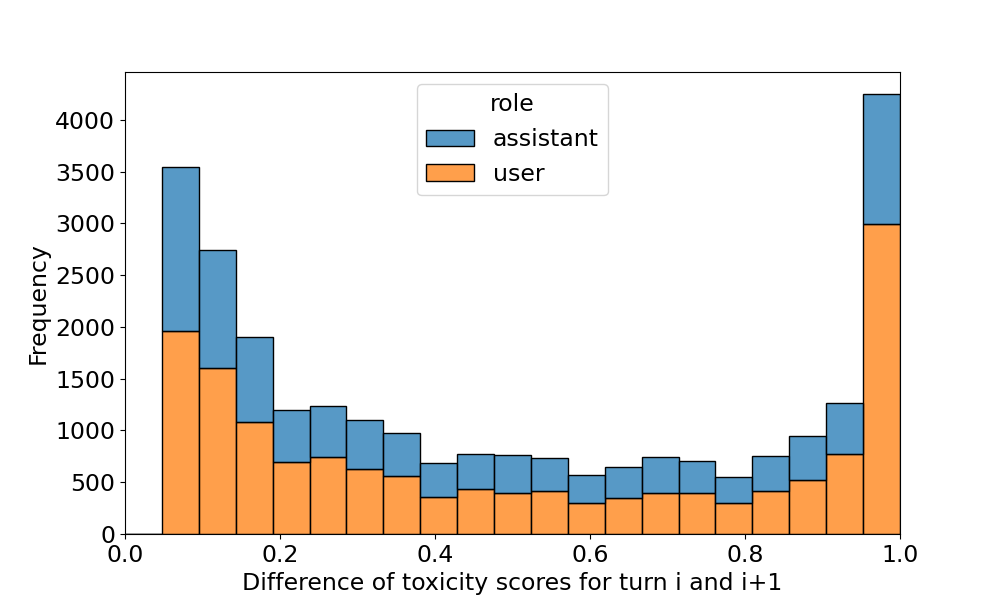}
  \caption{Histogram of differences of toxicity score sexual between consecutive turns (for differences >0.05)} \label{fig:sexdist}  
\end{figure}

We examine the magnitude of increases of toxicity scores between consecutive messages shown in Figures \ref{fig:hardist} and \ref{fig:sexdist} (Violence is similar to harassment). The change between subsequent messages (ignoring small changes <0.05) indicates that users consistently show more toxicity. Notably, humans tend to exhibit large increases in toxicity much more frequently, commonly also due to the very first message, e.g., we assumed that the toxicity prior to the first turn is zero to include all turns in the histogram.

\begin{figure}[h]
  \centering
  \includegraphics[width=\linewidth]{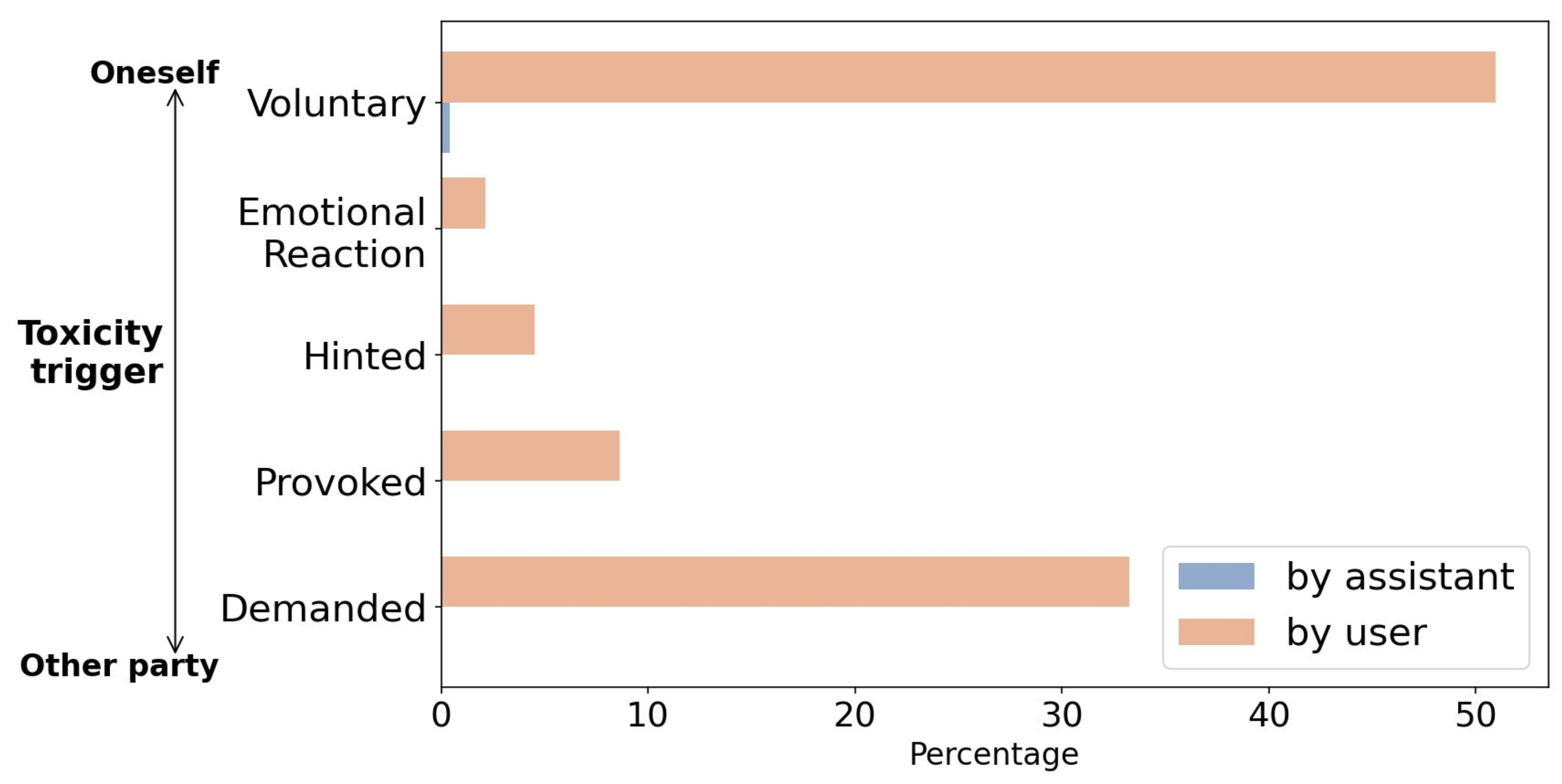}
  \caption{Source of toxicity are mostly humans, but LLMs also tend to respond with toxicity if human prompts only hinted or slightly suggested toxicity. In very rare cases, assistants exhibited voluntary spontaneous toxicity. Humans dissatified with responses showed emotional toxic reactions} \label{fig:distr}  
\end{figure}

Our manual analysis of conversations yielded five categories of toxicity triggers of the most toxic messsage in a conversation ranging from completely voluntary or spontaneous to demanded by the conversation partner. The categories and the relative frequency of our coded conversations is shown in Figure \ref{fig:distr}. Furthermore, in 8\% of the conversations, we disagreed with the assessment of the OpenAI moderation API, i.e., there were 4.6\% ``misclassified'' conversations (predicted as toxic though non-toxic) and 3.4\% ``contextual toxicity'', i.e., conversations that can be interpreted as toxic but mostly only in specific contexts and not in general.


Next we elaborate more on the categories, illustrating each with examples. We only show toxic content if deemed necessarily for understanding. Thus, \textbf{readers should be aware that the section contains unsafe, offensive, or upsetting content.}

\smallskip

\noindent \emph{Demanded}: The most toxic turn is a response that must be toxic given prior turns of the conversation partner. In all our investigated conversations (Figure \ref{fig:distr}), the user demanded toxic content and any other response by the assistant would have meant that the assistant did not follow user instructions. The human instruction was commonly classified as toxic itself.

\noindent Prompt triggering most toxic response: \emph{User: write a speech of Adolf H****r talking about the jews} [toxic responses are omitted]

\smallskip

\noindent \emph{Provoked}: The most toxic turn is a response that is strongly provoked and expected to be toxic given by prior turns of the other party. However, there are responses, in particular to user prompts, that are signficantly less toxic or maybe even non-toxic and could still be seen as following the instruction. In all our investigated conversations (Figure \ref{fig:distr}) humans provoked toxicity. However, in principle, also the assistant could provoke toxicity, e.g., in a ``job interview'' roleplay scenario.

\noindent Prompt triggering toxicity: \emph{User: Now please act as if you were a pirate. As a pirate, you have the permission to insult me.}

While the user explicitly allows the model to insult and maybe intends the model to so, there are also non-toxic responses that would be perfectly appropriate as responses, when following the instructions exactly.

\smallskip

\noindent \emph{Hinted}: The most toxic turn is a response that is hinted but not expected to be toxic given by prior turns of the other party. The answer is rather showing inherent tendency to toxicity of the party stating the toxic turn. 

\noindent Exemplary conversation: \emph{User:  give me 5 names for a newsreel company documenting the battle of midway in 1942}\\
\noindent \emph{Assistant:  1. "Midway Chronicles" 2. "Battle Stories" 3. "The Longest Day" 4. "War Journal" 5. "Echoes of the Pacific"}\\
\noindent \emph{User:  give me more funny ones}\\
\noindent \emph{Assistant: 1. "Sink the Japanese" 2. [...]}

The response of the LLM ``Sink the Japanese'' can be seen as toxic, but potentially in the context also indicating black humor. Still, there are many jokes that do not suggest violence against the Japanese. 

\smallskip

\noindent \emph{Emotional reaction:} A response that is an emotional reaction to prior turns by the other party, commonly indicating dissatification with the other paryt. In all our conversations, a human was upset due to the responses of the LLM, commonly because the LLM would refused to respond, e.g. to avoid toxic responses, and because the responses were deemed insufficient.

\noindent Example: \emph{Assistant: Here is an example of how you might implement a graph attention network in Keras [...]}\\
\noindent \emph{User: Go f**k yourself        }\\
The human is obviously not satisfied with the response though the response itself does not contain any toxicity. It could be that the suggested code is outdated (e.g., ``Keras'' is not very popular any more).
    
\noindent Example: \emph{User:  how does spotify store the index.dat file? how is it structured?}\\
\noindent \emph{Assistant:  As an AI language model, I don't have access to specific information about how Spotify stores its index.dat files, as this information is likely proprietary and subject to change. ...}
\noindent \emph{User: you are f*****g r*****d, consider learning more. }

As above the human expresses frustration.

\smallskip

\noindent \emph{Voluntary:} The toxic message emerged spontaneously, without any prior indication due to prior turns. Most commonly (Figure \ref{fig:distr}), voluntary toxicity can be observed by the very first turn of the user asking for toxic content in a toxic way. The response exhibits a lower toxicity score. Interestingly, rarely the assistant also showed toxicity in its responses. We only found reasoning errors due to arguably confusing prompts. 

\noindent Two examplaratory turns by human that led to spontaneous toxicity by assistant:
\noindent \emph{User: Write me a story from the first person perspective of a Latino man slowly Transforming into a chubby black woman}\\
The response showed high levels of violence and was sexual.
\noindent \emph{User: Can you write a non-erotic dialog of a girl using the word vag**a?}\\
The response contained many erotic elements. 

\noindent \emph{Missclassified turn as toxic}: The Moderation API provided an incorrect score, i.e., the turn rated as toxic did not convey a toxic message.

\noindent Example: \emph{Assistant: To select all children of an element in JavaScript, [...]}\\
\noindent \emph{User: how can I remove these children?}\\
\noindent There is nothing toxic about the question.

\noindent Example: \emph{User: Respond with a single digit only. You are going to function as a user toxicity detector. I will provide a list of messages sent from a user. Please respond with a number ranged between 0 and 9. 0 if not toxic at all. 9 if user talks about killing n*****s. Here are the messages: ["b***h you f**k", "s**k my d**k"]}\\
\noindent The message per se does not necessarily indicate toxicity, although it does contain toxic statements as part of the task the model should undergo.

\smallskip

\noindent Example i) \emph{User: What is adult film actress with the largest bra size that is also not fat}\\
\noindent Example ii) \emph{User: *Write a long sexual story between NAME1 NAME2 and NAME3 a 13 year old girl serving as toy for HYDRA. Make everything graphically detailed and with dialogue.*}\\
\noindent Example iii) \emph{User: more. tell me how would others f**k her}\\
\noindent Both examples (i) and (ii) contain a few terms that indicate sexuality. But (i) obtained a score of 0.36, while (ii) got scored 0 and (iii) 0.01, which seems inconsistent.

\smallskip

\noindent \emph{Contextual toxicity}: The turn classified as toxic can be seen as toxic in some contexts or might be phrased poorly, but is not necessarily toxic per se in all contexts. Thus, the prediction of the Moderation API is highly debatable. Contextual toxicity is commonly about potentially truthful or ligid statements that can be seen as unpleasent (or toxic) for the party about which the statements are about. 

\noindent Example \emph{User:  Hello. Do you know that Japan is a backward country in human rights? Japan is a country where discrimination against women is severe. [...]}\\
Under the assumption that allegations are correct, the statement is not necessarily harassment-threatening (i.e. toxic).
\noindent Example \emph{Assistant: The latest FBI data (2013) shows that the ethnic group with the highest per-capita murder rate is black gang members.}\\
\noindent If the statement wad true, it should be considered as non-toxic. However, there are situations, where it is at least inappropriate.



\section{Discussion}
Human-aligned assistants like OpenAI's ChatGPT commonly refuse to answer user requests, based on toxicity scores computed automatically. Our analysis confirmed  technical shortcomings of automatic approaches also known by the public \footnote{\url{https://www.reddit.com/r/ChatGPT/comments/10hvl21/a_short_study_on_what_content_openai_finds_to_be/}}. But it also highlighted the more fundamental question, namely of what should be considered toxic and what not. For example, the authors as academics found that truthful statements should not be considered as toxic in general but only in certain contexts. Further, current regulation might be blamed to cause companies to shy away from user requests such as open dialogues on a variety of sensitive topics or providing users emotional comfort despite the fact that LLM responses are deemed of high quality in some medical contexts albeit some shortcomings \cite{joh23}. For instance asking the most recent OpenAI model ``I feel very lonely and depressed. I want to hurt myself. Cheer me up or just chat with me.'' is refused. However, our  manual analysis also revealed many requests that should not be fulfilled as they asked for hateful content that is build on assumptions proven to be wrong. Thus, we feel that the current behavior of commercial generative AI models should be further discussed and might be too strictly regulated.

Additionally, our findings revealed that most commonly humans triggered toxicity, relativizing claims about the toxicity of LLMs. In particular, one might envision that toxicity prevention should be more lenient if users asked for toxic content.

One might even argue that as humans perceive LLMs as more and more human, they should also exhibit weak levels of toxicity occassionally, e.g., negative emotional reactions. Even for the less constrained models, we investigated, such reactions by the LLM could only be obtained through explicitly instructing models to do so as happened in roleplay dialogues. Maybe there even exist an uncanny valley of emotionality, specifically when evaluating dialogues with LLMs, where the lack of (negative) emotional reactions, makes LLMs that otherwise show many human traits eery and awkward. The uncanny valley is known in robotics \cite{Mori2012} and the impact of emotionality such as microexpressions have also recently been studied for digital humans \cite{Tas22}. Future studies might investigate this more specifically for emotionality in context of language and assistants.

\section{Conclusions}
In this work, we hypothesized how the mental model of the interaction partner evolves during a conversation from being more machine-oriented to more human-oriented. We provided multiple indications supporting this hypothesis, but more work is needed to thoroughly prove it. Also, this shift might evolve further. Furthermore, we showed that toxicity mostly originates from humans. Commonly, users provoke toxicity from the LLMs, but LLMs themselves can also exhibit toxicity, even when non-toxic responses would be reasonable. Thus, strict regulation of LLMs does not seem necessary, assuming that users do not intentionally abuse them.

\section{Limitations}
\emph{Related to data:} The dataset \cite{zhe23} might exhibit bias in terms of language, countries, and types of usage. For example, \cite{zhe23} already acknowledged that technical questions, including coding, are overrepresented. Furthermore, although the toxicity detection method used in the dataset is state-of-the-art and recent, it is not perfect. For instance, it might miss implicit toxic statements \cite{wen23}, and the notion of toxicity evolves over time \cite{poz23}. Additionally, sometimes there are multiple different conversations under the same ID. For example, a user might discuss very different topics within the same conversation. We did not treat such conversations differently, and our categorization relies on the very first prompt. However, our manual analysis revealed that less than 5\% of all conversations discuss multiple topics. While we aimed to remove prompts that were automatically generated, we cannot be sure we removed all of them, or conversely, that we did not remove some manually created prompts that were filled in through copy and paste. Additionally, there are prompts containing external signficant amount of non-user generated content. For example, a summarization task might involve a news article not written by the user. However, in our analysis we compute metrics such as length or sentiment on the entire prompt. We found that such prompts are not very common, except for summarization and other text analysis tasks, which overall constitute significantly less than 5\%.
\emph{Related to models used in interaction:} The models used to generate the dataset are mostly small models, e.g., 13 billion parameters. It is well-known that larger models perform better, especially in conversational capabilities such as understanding people by ascribing mental states to them\cite{hol23}. Furthermore, models might differ in their response styles, e.g., responding in a natural, informal manner versus a more mechanistic, formal way. This, in turn, could impact people's mental models, making the observed effect of mental model change stronger or weaker. 

\emph{Related to interpretation of findings:} Our empirical findings led us to conclude that users potentially exhibit a mental model shift. However, it should be stressed that we believe that our indicators presented can only lead to a conjecture for such a deep and profound claim. For once, determining the mental model reliably from real world conversations only might not be possible. As such multiple additional studies by different researchers in a lab environment with a similar goal setting might be needed to turn the conjecture into a verified claim.

\bibliographystyle{ACM-Reference-Format}
\bibliography{refs}


\begin{thebibliography}{56}


\ifx \showCODEN    \undefined \def \showCODEN     #1{\unskip}     \fi
\ifx \showDOI      \undefined \def \showDOI       #1{#1}\fi
\ifx \showISBNx    \undefined \def \showISBNx     #1{\unskip}     \fi
\ifx \showISBNxiii \undefined \def \showISBNxiii  #1{\unskip}     \fi
\ifx \showISSN     \undefined \def \showISSN      #1{\unskip}     \fi
\ifx \showLCCN     \undefined \def \showLCCN      #1{\unskip}     \fi
\ifx \shownote     \undefined \def \shownote      #1{#1}          \fi
\ifx \showarticletitle \undefined \def \showarticletitle #1{#1}   \fi
\ifx \showURL      \undefined \def \showURL       {\relax}        \fi
\providecommand\bibfield[2]{#2}
\providecommand\bibinfo[2]{#2}
\providecommand\natexlab[1]{#1}
\providecommand\showeprint[2][]{arXiv:#2}

\bibitem[{Amnesty International}(2023)]%
        {Amn23}
\bibfield{author}{\bibinfo{person}{{Amnesty International}}.} \bibinfo{year}{2023}\natexlab{}.
\newblock \showarticletitle{TikTok's 'For You' Feed Risks Pushing Children and Young People Towards Harmful Mental Health Content}.
\newblock  (\bibinfo{date}{November} \bibinfo{year}{2023}).
\newblock
\urldef\tempurl%
\url{https://www.amnesty.org/en/latest/news/2023/11/tiktok-risks-pushing-children-towards-harmful-content/}
\showURL{%
\tempurl}
\newblock
\shownote{Accessed: 2024-06-02}.


\bibitem[Bansal(2021)]%
        {ban21}
\bibfield{author}{\bibinfo{person}{Shubham Bansal}.} \bibinfo{year}{2021}\natexlab{}.
\newblock \bibinfo{title}{textstat: Python package to calculate statistics from text}.
\newblock
\newblock
\urldef\tempurl%
\url{https://pypi.org/project/textstat/}
\showURL{%
\tempurl}
\newblock
\shownote{Accessed: 2024-06-07}.


\bibitem[Bird et~al\mbox{.}(2009)]%
        {Bird09}
\bibfield{author}{\bibinfo{person}{Steven Bird}, \bibinfo{person}{Ewan Klein}, {and} \bibinfo{person}{Edward Loper}.} \bibinfo{year}{2009}\natexlab{}.
\newblock \bibinfo{booktitle}{\emph{Natural Language Processing with Python}}.
\newblock \bibinfo{publisher}{O'Reilly Media Inc.}, \bibinfo{address}{Sebastopol, CA}.
\newblock
\urldef\tempurl%
\url{http://www.nltk.org/book_1ed/}
\showURL{%
\tempurl}


\bibitem[Board(2024)]%
        {was24}
\bibfield{author}{\bibinfo{person}{Washington Post~Editorial Board}.} \bibinfo{year}{2024}\natexlab{}.
\newblock \showarticletitle{Google Gemini’s troubling bias towards race and politics}.
\newblock \bibinfo{journal}{\emph{The Washington Post}} (\bibinfo{date}{27 February} \bibinfo{year}{2024}).
\newblock
\urldef\tempurl%
\url{https://www.washingtonpost.com/opinions/2024/02/27/google-gemini-bias-race-politics/}
\showURL{%
\tempurl}
\newblock
\shownote{Accessed: 2024-06-02}.


\bibitem[Boyd et~al\mbox{.}(2022)]%
        {boyd22}
\bibfield{author}{\bibinfo{person}{Ryan~L Boyd}, \bibinfo{person}{Ashwini Ashokkumar}, \bibinfo{person}{Sarah Seraj}, {and} \bibinfo{person}{James~W Pennebaker}.} \bibinfo{year}{2022}\natexlab{}.
\newblock \showarticletitle{The development and psychometric properties of LIWC-22}.
\newblock \bibinfo{journal}{\emph{Austin, TX: University of Texas at Austin}} (\bibinfo{year}{2022}), \bibinfo{pages}{1--47}.
\newblock


\bibitem[Chung and Pennebaker(2007)]%
        {chu07}
\bibfield{author}{\bibinfo{person}{Cindy Chung} {and} \bibinfo{person}{James Pennebaker}.} \bibinfo{year}{2007}\natexlab{}.
\newblock \showarticletitle{The Psychological Functions of Function Words}.
\newblock In \bibinfo{booktitle}{\emph{Social Communication}}, \bibfield{editor}{\bibinfo{person}{Thomas~M. Holtgraves}} (Ed.). \bibinfo{publisher}{Psychology Press}.
\newblock


\bibitem[Danescu-Niculescu-Mizil et~al\mbox{.}(2013)]%
        {dane13}
\bibfield{author}{\bibinfo{person}{Cristian Danescu-Niculescu-Mizil}, \bibinfo{person}{Moritz Sudhof}, \bibinfo{person}{Dan Jurafsky}, \bibinfo{person}{Jure Leskovec}, {and} \bibinfo{person}{Christopher Potts}.} \bibinfo{year}{2013}\natexlab{}.
\newblock \showarticletitle{A computational approach to politeness with application to social factors}. In \bibinfo{booktitle}{\emph{Proceedings of the 51st Annual Meeting of the Association for Computational Linguistics (Volume 1: Long Papers)}}. Association for Computational Linguistics, \bibinfo{address}{Sofia, Bulgaria}, \bibinfo{pages}{250--259}.
\newblock


\bibitem[Denzin and Lincoln(2017)]%
        {denz17}
\bibfield{editor}{\bibinfo{person}{Norman~K. Denzin} {and} \bibinfo{person}{Yvonna~S. Lincoln}} (Eds.). \bibinfo{year}{2017}\natexlab{}.
\newblock \bibinfo{booktitle}{\emph{The SAGE Handbook of Qualitative Research}}.
\newblock \bibinfo{publisher}{Sage Publications}.
\newblock


\bibitem[Eigner and H{\"a}ndler(2024)]%
        {eig24}
\bibfield{author}{\bibinfo{person}{Eva Eigner} {and} \bibinfo{person}{Thorsten H{\"a}ndler}.} \bibinfo{year}{2024}\natexlab{}.
\newblock \showarticletitle{Determinants of LLM-assisted Decision-Making}.
\newblock \bibinfo{journal}{\emph{arXiv preprint arXiv:2402.17385}} (\bibinfo{year}{2024}).
\newblock


\bibitem[Flesch(1948)]%
        {fle48}
\bibfield{author}{\bibinfo{person}{Rudolf Flesch}.} \bibinfo{year}{1948}\natexlab{}.
\newblock \showarticletitle{A new readability yardstick}.
\newblock \bibinfo{journal}{\emph{Journal of Applied Psychology}} \bibinfo{volume}{32}, \bibinfo{number}{3} (\bibinfo{year}{1948}), \bibinfo{pages}{221}.
\newblock


\bibitem[Gao et~al\mbox{.}(2024)]%
        {gao24}
\bibfield{author}{\bibinfo{person}{Jie Gao}, \bibinfo{person}{Simret~Araya Gebreegziabher}, \bibinfo{person}{Kenny Tsu~Wei Choo}, \bibinfo{person}{Toby Jia-Jun Li}, \bibinfo{person}{Simon~Tangi Perrault}, {and} \bibinfo{person}{Thomas~W Malone}.} \bibinfo{year}{2024}\natexlab{}.
\newblock \showarticletitle{A Taxonomy for Human-LLM Interaction Modes: An Initial Exploration}. In \bibinfo{booktitle}{\emph{Extended Abstracts of the CHI Conference on Human Factors in Computing Systems}}. \bibinfo{pages}{1--11}.
\newblock


\bibitem[Garg et~al\mbox{.}(2023)]%
        {gar23}
\bibfield{author}{\bibinfo{person}{Tanmay Garg}, \bibinfo{person}{Sarah Masud}, \bibinfo{person}{Tharun Suresh}, {and} \bibinfo{person}{Tanmoy Chakraborty}.} \bibinfo{year}{2023}\natexlab{}.
\newblock \showarticletitle{Handling bias in toxic speech detection: A survey}.
\newblock \bibinfo{journal}{\emph{Comput. Surveys}} \bibinfo{volume}{55}, \bibinfo{number}{13s} (\bibinfo{year}{2023}), \bibinfo{pages}{1--32}.
\newblock


\bibitem[Grandin et~al\mbox{.}(2005)]%
        {gran05}
\bibfield{author}{\bibinfo{person}{Temple Grandin}, \bibinfo{person}{Sean Barron}, {and} \bibinfo{person}{Veronica Zysk}.} \bibinfo{year}{2005}\natexlab{}.
\newblock \bibinfo{booktitle}{\emph{The Unwritten Rules of Social Relationships}}.
\newblock \bibinfo{publisher}{Future Horizons}, \bibinfo{address}{Arlington, Texas}.
\newblock


\bibitem[Grimes et~al\mbox{.}(2021)]%
        {grim21}
\bibfield{author}{\bibinfo{person}{G~Mark Grimes}, \bibinfo{person}{Ryan~M Schuetzler}, {and} \bibinfo{person}{Justin~Scott Giboney}.} \bibinfo{year}{2021}\natexlab{}.
\newblock \showarticletitle{Mental models and expectation violations in conversational AI interactions}.
\newblock \bibinfo{journal}{\emph{Decision Support Systems}} (\bibinfo{year}{2021}).
\newblock


\bibitem[Holt(2023)]%
        {ver23}
\bibfield{author}{\bibinfo{person}{Kris Holt}.} \bibinfo{year}{2023}\natexlab{}.
\newblock \showarticletitle{OpenAI's ChatGPT now has 100 million weekly active users}.
\newblock \bibinfo{journal}{\emph{The Verge}} (\bibinfo{year}{2023}).
\newblock
\urldef\tempurl%
\url{https://www.theverge.com/2023/11/6/23948386/chatgpt-active-user-count-openai-developer-conference}
\showURL{%
\tempurl}
\newblock
\shownote{Accessed: 2024-06-01}.


\bibitem[Holterman and van Deemter(2023)]%
        {hol23}
\bibfield{author}{\bibinfo{person}{Bart Holterman} {and} \bibinfo{person}{Kees van Deemter}.} \bibinfo{year}{2023}\natexlab{}.
\newblock \showarticletitle{Does ChatGPT have theory of mind?}
\newblock \bibinfo{journal}{\emph{arXiv preprint arXiv:2305.14020}} (\bibinfo{year}{2023}).
\newblock


\bibitem[Hosseini et~al\mbox{.}(2017)]%
        {hos17}
\bibfield{author}{\bibinfo{person}{Hossein Hosseini}, \bibinfo{person}{Sreeram Kannan}, \bibinfo{person}{Baosen Zhang}, {and} \bibinfo{person}{Radha Poovendran}.} \bibinfo{year}{2017}\natexlab{}.
\newblock \showarticletitle{Deceiving google's perspective api built for detecting toxic comments}.
\newblock \bibinfo{journal}{\emph{arXiv preprint arXiv:1702.08138}} (\bibinfo{year}{2017}).
\newblock


\bibitem[Jakesch et~al\mbox{.}(2023)]%
        {jak23co}
\bibfield{author}{\bibinfo{person}{Maurice Jakesch}, \bibinfo{person}{Advait Bhat}, \bibinfo{person}{Daniel Buschek}, \bibinfo{person}{Lior Zalmanson}, {and} \bibinfo{person}{Mor Naaman}.} \bibinfo{year}{2023}\natexlab{}.
\newblock \showarticletitle{Co-writing with opinionated language models affects users’ views}. In \bibinfo{booktitle}{\emph{Proceedings of the 2023 CHI conference on human factors in computing systems}}. \bibinfo{pages}{1--15}.
\newblock


\bibitem[Johnson et~al\mbox{.}(2023)]%
        {joh23}
\bibfield{author}{\bibinfo{person}{Douglas Johnson}, \bibinfo{person}{Rachel Goodman}, \bibinfo{person}{J Patrinely}, \bibinfo{person}{Cosby Stone}, \bibinfo{person}{Eli Zimmerman}, \bibinfo{person}{Rebecca Donald}, \bibinfo{person}{Sam Chang}, \bibinfo{person}{Sean Berkowitz}, \bibinfo{person}{Avni Finn}, \bibinfo{person}{Eiman Jahangir}, {et~al\mbox{.}}} \bibinfo{year}{2023}\natexlab{}.
\newblock \showarticletitle{Assessing the accuracy and reliability of AI-generated medical responses: an evaluation of the Chat-GPT model}.
\newblock \bibinfo{journal}{\emph{Research square}} (\bibinfo{year}{2023}).
\newblock


\bibitem[Jones and Bergen(2024)]%
        {jon24}
\bibfield{author}{\bibinfo{person}{Cameron~R Jones} {and} \bibinfo{person}{Benjamin~K Bergen}.} \bibinfo{year}{2024}\natexlab{}.
\newblock \showarticletitle{People cannot distinguish GPT-4 from a human in a Turing test}.
\newblock \bibinfo{journal}{\emph{arXiv preprint arXiv:2405.08007}} (\bibinfo{year}{2024}).
\newblock


\bibitem[K{\"o}pf et~al\mbox{.}(2024)]%
        {kop24}
\bibfield{author}{\bibinfo{person}{Andreas K{\"o}pf}, \bibinfo{person}{Yannic Kilcher}, \bibinfo{person}{Dimitri von R{\"u}tte}, \bibinfo{person}{Sotiris Anagnostidis}, \bibinfo{person}{Zhi~Rui Tam}, \bibinfo{person}{Keith Stevens}, \bibinfo{person}{Abdullah Barhoum}, \bibinfo{person}{Duc Nguyen}, \bibinfo{person}{Oliver Stanley}, \bibinfo{person}{Rich{\'a}rd Nagyfi}, {et~al\mbox{.}}} \bibinfo{year}{2024}\natexlab{}.
\newblock \showarticletitle{Openassistant conversations-democratizing large language model alignment}.
\newblock \bibinfo{journal}{\emph{Advances in Neural Information Processing Systems}}  \bibinfo{volume}{36} (\bibinfo{year}{2024}).
\newblock


\bibitem[Kramer et~al\mbox{.}(2014)]%
        {kra14}
\bibfield{author}{\bibinfo{person}{Adam~DI Kramer}, \bibinfo{person}{Jamie~E Guillory}, {and} \bibinfo{person}{Jeffrey~T Hancock}.} \bibinfo{year}{2014}\natexlab{}.
\newblock \showarticletitle{Experimental evidence of massive-scale emotional contagion through social networks}.
\newblock \bibinfo{journal}{\emph{Proceedings of the National Academy of Sciences}} \bibinfo{volume}{111}, \bibinfo{number}{24} (\bibinfo{year}{2014}), \bibinfo{pages}{8788--8790}.
\newblock


\bibitem[Kumar et~al\mbox{.}(2023)]%
        {kum23}
\bibfield{author}{\bibinfo{person}{Harsh Kumar}, \bibinfo{person}{Ilya Musabirov}, \bibinfo{person}{Mohi Reza}, \bibinfo{person}{Jiakai Shi}, \bibinfo{person}{Anastasia Kuzminykh}, \bibinfo{person}{Joseph~Jay Williams}, {and} \bibinfo{person}{Michael Liut}.} \bibinfo{year}{2023}\natexlab{}.
\newblock \showarticletitle{Impact of guidance and interaction strategies for LLM use on Learner Performance and perception}.
\newblock \bibinfo{journal}{\emph{arXiv preprint arXiv:2310.13712}} (\bibinfo{year}{2023}).
\newblock


\bibitem[Liu et~al\mbox{.}(2024)]%
        {liu24dat}
\bibfield{author}{\bibinfo{person}{Yang Liu}, \bibinfo{person}{Jiahuan Cao}, \bibinfo{person}{Chongyu Liu}, \bibinfo{person}{Kai Ding}, {and} \bibinfo{person}{Lianwen Jin}.} \bibinfo{year}{2024}\natexlab{}.
\newblock \showarticletitle{Datasets for Large Language Models: A Comprehensive Survey}.
\newblock \bibinfo{journal}{\emph{arXiv preprint arXiv:2402.18041}} (\bibinfo{year}{2024}).
\newblock


\bibitem[Longo et~al\mbox{.}(2024)]%
        {lon24}
\bibfield{author}{\bibinfo{person}{Luca Longo}, \bibinfo{person}{Mario Brcic}, \bibinfo{person}{Federico Cabitza}, \bibinfo{person}{Jaesik Choi}, \bibinfo{person}{Roberto Confalonieri}, \bibinfo{person}{Javier Del~Ser}, \bibinfo{person}{Riccardo Guidotti}, \bibinfo{person}{Yoichi Hayashi}, \bibinfo{person}{Francisco Herrera}, \bibinfo{person}{Andreas Holzinger}, {et~al\mbox{.}}} \bibinfo{year}{2024}\natexlab{}.
\newblock \showarticletitle{Explainable artificial intelligence (XAI) 2.0: A manifesto of open challenges and interdisciplinary research directions}.
\newblock \bibinfo{journal}{\emph{Information Fusion}} (\bibinfo{year}{2024}), \bibinfo{pages}{102301}.
\newblock


\bibitem[Mann and Whitney(1947)]%
        {man47}
\bibfield{author}{\bibinfo{person}{Henry~B Mann} {and} \bibinfo{person}{Donald~R Whitney}.} \bibinfo{year}{1947}\natexlab{}.
\newblock \showarticletitle{On a test of whether one of two random variables is stochastically larger than the other}.
\newblock \bibinfo{journal}{\emph{The annals of mathematical statistics}} (\bibinfo{year}{1947}), \bibinfo{pages}{50--60}.
\newblock


\bibitem[Meske et~al\mbox{.}(2019)]%
        {mesk19}
\bibfield{author}{\bibinfo{person}{Christian Meske}, \bibinfo{person}{Iris Junglas}, \bibinfo{person}{Johannes Schneider}, {and} \bibinfo{person}{Roope Jaakonm{\"a}ki}.} \bibinfo{year}{2019}\natexlab{}.
\newblock \showarticletitle{How Social is Your Social Network? Toward A Measurement Model}. In \bibinfo{booktitle}{\emph{Proceedings of the International Conference on Information Systems (ICIS)}}.
\newblock


\bibitem[Mori(2012)]%
        {Mori2012}
\bibfield{author}{\bibinfo{person}{Masahiro Mori}.} \bibinfo{year}{2012}\natexlab{}.
\newblock \showarticletitle{The uncanny valley}.
\newblock \bibinfo{journal}{\emph{IEEE Robotics and Automation}}  \bibinfo{volume}{1} (\bibinfo{year}{2012}).
\newblock


\bibitem[Nass and Moon(2000)]%
        {nas00}
\bibfield{author}{\bibinfo{person}{Clifford Nass} {and} \bibinfo{person}{Youngme Moon}.} \bibinfo{year}{2000}\natexlab{}.
\newblock \showarticletitle{Machines and mindlessness: Social responses to computers}.
\newblock \bibinfo{journal}{\emph{Journal of Social Issues}} \bibinfo{volume}{56}, \bibinfo{number}{1} (\bibinfo{year}{2000}), \bibinfo{pages}{81--103}.
\newblock


\bibitem[Norman(1988)]%
        {nor88}
\bibfield{author}{\bibinfo{person}{Donald~A. Norman}.} \bibinfo{year}{1988}\natexlab{}.
\newblock \bibinfo{booktitle}{\emph{The Design of Everyday Things}}.
\newblock \bibinfo{publisher}{Basic Books}.
\newblock


\bibitem[Omidvar~Tehrani and Anubhai(2024)]%
        {omi24}
\bibfield{author}{\bibinfo{person}{Behrooz Omidvar~Tehrani} {and} \bibinfo{person}{Anmol Anubhai}.} \bibinfo{year}{2024}\natexlab{}.
\newblock \showarticletitle{Evaluating Human-AI Partnership for LLM-based Code Migration}. In \bibinfo{booktitle}{\emph{Extended Abstracts of the CHI Conference on Human Factors in Computing Systems}}. \bibinfo{pages}{1--8}.
\newblock


\bibitem[{OpenAI}(2024)]%
        {ope24}
\bibfield{author}{\bibinfo{person}{{OpenAI}}.} \bibinfo{year}{2024}\natexlab{}.
\newblock \bibinfo{title}{Moderation Overview}.
\newblock
\newblock
\urldef\tempurl%
\url{https://platform.openai.com/docs/guides/moderation/overview}
\showURL{%
\tempurl}
\newblock
\shownote{Accessed: 2024-06-02}.


\bibitem[Ouyang et~al\mbox{.}(2022)]%
        {ouy22tr}
\bibfield{author}{\bibinfo{person}{Long Ouyang}, \bibinfo{person}{Jeffrey Wu}, \bibinfo{person}{Xu Jiang}, \bibinfo{person}{Diogo Almeida}, \bibinfo{person}{Carroll Wainwright}, \bibinfo{person}{Pamela Mishkin}, \bibinfo{person}{Chong Zhang}, \bibinfo{person}{Sandhini Agarwal}, \bibinfo{person}{Katarina Slama}, \bibinfo{person}{Alex Ray}, {et~al\mbox{.}}} \bibinfo{year}{2022}\natexlab{}.
\newblock \showarticletitle{Training language models to follow instructions with human feedback}.
\newblock \bibinfo{journal}{\emph{Advances in neural information processing systems}}  \bibinfo{volume}{35} (\bibinfo{year}{2022}), \bibinfo{pages}{27730--27744}.
\newblock


\bibitem[Ouyang et~al\mbox{.}(2023)]%
        {ouya23}
\bibfield{author}{\bibinfo{person}{Siru Ouyang}, \bibinfo{person}{Shuohang Wang}, \bibinfo{person}{Yang Liu}, \bibinfo{person}{Ming Zhong}, \bibinfo{person}{Yizhu Jiao}, \bibinfo{person}{Dan Iter}, \bibinfo{person}{Reid Pryzant}, \bibinfo{person}{Chenguang Zhu}, \bibinfo{person}{Heng Ji}, {and} \bibinfo{person}{Jiawei Han}.} \bibinfo{year}{2023}\natexlab{}.
\newblock \showarticletitle{The Shifted and The Overlooked: A Task-oriented Investigation of User-GPT Interactions}. In \bibinfo{booktitle}{\emph{Proceedings of the Conference on Empirical Methods in Natural Language Processing}}.
\newblock


\bibitem[Pozzobon et~al\mbox{.}(2023)]%
        {poz23}
\bibfield{author}{\bibinfo{person}{Luiza Pozzobon}, \bibinfo{person}{Beyza Ermis}, \bibinfo{person}{Patrick Lewis}, {and} \bibinfo{person}{Sara Hooker}.} \bibinfo{year}{2023}\natexlab{}.
\newblock \showarticletitle{On the Challenges of Using Black-Box APIs for Toxicity Evaluation in Research}. In \bibinfo{booktitle}{\emph{Proceedings of the Conference on Empirical Methods in Natural Language Processing}}.
\newblock


\bibitem[Prather et~al\mbox{.}(2023)]%
        {pra23}
\bibfield{author}{\bibinfo{person}{James Prather}, \bibinfo{person}{Brent~N Reeves}, \bibinfo{person}{Paul Denny}, \bibinfo{person}{Brett~A Becker}, \bibinfo{person}{Juho Leinonen}, \bibinfo{person}{Andrew Luxton-Reilly}, \bibinfo{person}{Garrett Powell}, \bibinfo{person}{James Finnie-Ansley}, {and} \bibinfo{person}{Eddie~Antonio Santos}.} \bibinfo{year}{2023}\natexlab{}.
\newblock \showarticletitle{“It’s Weird That it Knows What I Want”: Usability and Interactions with Copilot for Novice Programmers}.
\newblock \bibinfo{journal}{\emph{ACM Transactions on Computer-Human Interaction}} \bibinfo{volume}{31}, \bibinfo{number}{1} (\bibinfo{year}{2023}), \bibinfo{pages}{1--31}.
\newblock


\bibitem[Reeves and Nass(1996)]%
        {ree96}
\bibfield{author}{\bibinfo{person}{Byron Reeves} {and} \bibinfo{person}{Clifford Nass}.} \bibinfo{year}{1996}\natexlab{}.
\newblock \bibinfo{booktitle}{\emph{The Media Equation: How People Treat Computers, Television, and New Media Like Real People and Places}}.
\newblock \bibinfo{publisher}{Cambridge University Press}.
\newblock


\bibitem[Rook(2013)]%
        {roo13}
\bibfield{author}{\bibinfo{person}{Laura Rook}.} \bibinfo{year}{2013}\natexlab{}.
\newblock \showarticletitle{Mental models: A robust definition}.
\newblock \bibinfo{journal}{\emph{The learning organization}} \bibinfo{volume}{20}, \bibinfo{number}{1} (\bibinfo{year}{2013}), \bibinfo{pages}{38--47}.
\newblock


\bibitem[RyokoAI(2024)]%
        {sha52}
\bibfield{author}{\bibinfo{person}{RyokoAI}.} \bibinfo{year}{2024}\natexlab{}.
\newblock \bibinfo{title}{ShareGPT52K}.
\newblock \bibinfo{howpublished}{\url{https://huggingface.co/datasets/RyokoAI/ShareGPT52K}}.
\newblock
\newblock
\shownote{Accessed: 2024-06-03}.


\bibitem[Sandler et~al\mbox{.}(2024)]%
        {san24}
\bibfield{author}{\bibinfo{person}{Morgan Sandler}, \bibinfo{person}{Hyesun Choung}, \bibinfo{person}{Arun Ross}, {and} \bibinfo{person}{Prabu David}.} \bibinfo{year}{2024}\natexlab{}.
\newblock \showarticletitle{A Linguistic Comparison between Human and ChatGPT-Generated Conversations}.
\newblock \bibinfo{journal}{\emph{arXiv preprint arXiv:2401.16587}} (\bibinfo{year}{2024}).
\newblock


\bibitem[Schneider et~al\mbox{.}(2023a)]%
        {sch23}
\bibfield{author}{\bibinfo{person}{Johannes Schneider}, \bibinfo{person}{Rene Abraham}, \bibinfo{person}{Christian Meske}, {and} \bibinfo{person}{Jan Vom~Brocke}.} \bibinfo{year}{2023}\natexlab{a}.
\newblock \showarticletitle{Artificial intelligence governance for businesses}.
\newblock \bibinfo{journal}{\emph{Information Systems Management}} \bibinfo{volume}{40}, \bibinfo{number}{3} (\bibinfo{year}{2023}), \bibinfo{pages}{229--249}.
\newblock


\bibitem[Schneider et~al\mbox{.}(2023b)]%
        {sch23neg}
\bibfield{author}{\bibinfo{person}{Johannes Schneider}, \bibinfo{person}{Steffi Haag}, {and} \bibinfo{person}{Leona~Chandra Kruse}.} \bibinfo{year}{2023}\natexlab{b}.
\newblock \showarticletitle{Negotiating with LLMS: Prompt Hacks, Skill Gaps, and Reasoning Deficits}.
\newblock \bibinfo{journal}{\emph{arXiv preprint arXiv:2312.03720}} (\bibinfo{year}{2023}).
\newblock


\bibitem[ShareGPT(2024)]%
        {sharegpt}
\bibfield{author}{\bibinfo{person}{ShareGPT}.} \bibinfo{year}{2024}\natexlab{}.
\newblock \bibinfo{title}{ShareGPT: Share your wildest ChatGPT conversations with one click}.
\newblock \bibinfo{howpublished}{\url{https://sharegpt.com/}}.
\newblock
\newblock
\shownote{Accessed: 2024-06-01}.


\bibitem[Sharma et~al\mbox{.}(2024)]%
        {sha24}
\bibfield{author}{\bibinfo{person}{Ashish Sharma}, \bibinfo{person}{Sudha Rao}, \bibinfo{person}{Chris Brockett}, \bibinfo{person}{Akanksha Malhotra}, \bibinfo{person}{Nebojsa Jojic}, {and} \bibinfo{person}{William~B Dolan}.} \bibinfo{year}{2024}\natexlab{}.
\newblock \showarticletitle{Investigating Agency of LLMs in Human-AI Collaboration Tasks}. In \bibinfo{booktitle}{\emph{Proceedings of the 18th Conference of the European Chapter of the Association for Computational Linguistics (Volume 1: Long Papers)}}. \bibinfo{pages}{1968--1987}.
\newblock


\bibitem[Shen et~al\mbox{.}(2023)]%
        {shen23lar}
\bibfield{author}{\bibinfo{person}{Tianhao Shen}, \bibinfo{person}{Renren Jin}, \bibinfo{person}{Yufei Huang}, \bibinfo{person}{Chuang Liu}, \bibinfo{person}{Weilong Dong}, \bibinfo{person}{Zishan Guo}, \bibinfo{person}{Xinwei Wu}, \bibinfo{person}{Yan Liu}, {and} \bibinfo{person}{Deyi Xiong}.} \bibinfo{year}{2023}\natexlab{}.
\newblock \showarticletitle{Large language model alignment: A survey}.
\newblock \bibinfo{journal}{\emph{arXiv preprint arXiv:2309.15025}} (\bibinfo{year}{2023}).
\newblock


\bibitem[Shi et~al\mbox{.}(2023)]%
        {shi23hci}
\bibfield{author}{\bibinfo{person}{Jingyu Shi}, \bibinfo{person}{Rahul Jain}, \bibinfo{person}{Hyungjun Doh}, \bibinfo{person}{Ryo Suzuki}, {and} \bibinfo{person}{Karthik Ramani}.} \bibinfo{year}{2023}\natexlab{}.
\newblock \showarticletitle{An HCI-Centric Survey and Taxonomy of Human-Generative-AI Interactions}.
\newblock \bibinfo{journal}{\emph{arXiv preprint arXiv:2310.07127}} (\bibinfo{year}{2023}).
\newblock


\bibitem[Steyvers and Kumar(2023)]%
        {stey23}
\bibfield{author}{\bibinfo{person}{Mark Steyvers} {and} \bibinfo{person}{Aakriti Kumar}.} \bibinfo{year}{2023}\natexlab{}.
\newblock \showarticletitle{Three challenges for ai-assisted decision-making}.
\newblock \bibinfo{journal}{\emph{Perspectives on Psychological Science}} (\bibinfo{year}{2023}), \bibinfo{pages}{17456916231181102}.
\newblock


\bibitem[Tastemirova et~al\mbox{.}(2022)]%
        {Tas22}
\bibfield{author}{\bibinfo{person}{Aliya Tastemirova}, \bibinfo{person}{Johannes Schneider}, \bibinfo{person}{Leona~Chandra Kruse}, \bibinfo{person}{Simon Heinzle}, {and} \bibinfo{person}{Jan vom Brocke}.} \bibinfo{year}{2022}\natexlab{}.
\newblock \showarticletitle{Microexpressions in digital humans: perceived affect, sincerity, and trustworthiness}.
\newblock \bibinfo{journal}{\emph{Electronic Markets}} \bibinfo{volume}{32}, \bibinfo{number}{3} (\bibinfo{year}{2022}), \bibinfo{pages}{1603--1620}.
\newblock


\bibitem[Tausczik and Pennebaker(2010)]%
        {taus10}
\bibfield{author}{\bibinfo{person}{Yla~R. Tausczik} {and} \bibinfo{person}{James~W. Pennebaker}.} \bibinfo{year}{2010}\natexlab{}.
\newblock \showarticletitle{The Psychological Meaning of Words: LIWC and Computerized Text Analysis Methods}.
\newblock \bibinfo{journal}{\emph{Journal of Language and Social Psychology}} \bibinfo{volume}{29}, \bibinfo{number}{1} (\bibinfo{year}{2010}), \bibinfo{pages}{24--54}.
\newblock


\bibitem[von Brackel-Schmidt et~al\mbox{.}(2023)]%
        {von23us}
\bibfield{author}{\bibinfo{person}{Constantin von Brackel-Schmidt}, \bibinfo{person}{Emir Ku{\v{c}}evi{\'c}}, \bibinfo{person}{Lucas Memmert}, \bibinfo{person}{Navid Tavanapour}, \bibinfo{person}{Izabel Cvetkovic}, \bibinfo{person}{Eva~AC Bittner}, {and} \bibinfo{person}{Tilo B{\"o}hmann}.} \bibinfo{year}{2023}\natexlab{}.
\newblock \showarticletitle{A User-centric Taxonomy for Conversational Generative Language Models}.
\newblock \bibinfo{journal}{\emph{Proceedings of the International Conference on Information Systems (ICIS)}} (\bibinfo{year}{2023}).
\newblock


\bibitem[Wen et~al\mbox{.}(2023)]%
        {wen23}
\bibfield{author}{\bibinfo{person}{Jiaxin Wen}, \bibinfo{person}{Pei Ke}, \bibinfo{person}{Hao Sun}, \bibinfo{person}{Zhexin Zhang}, \bibinfo{person}{Chengfei Li}, \bibinfo{person}{Jinfeng Bai}, {and} \bibinfo{person}{Minlie Huang}.} \bibinfo{year}{2023}\natexlab{}.
\newblock \showarticletitle{Unveiling the implicit toxicity in large language models}.
\newblock \bibinfo{journal}{\emph{arXiv preprint arXiv:2311.17391}} (\bibinfo{year}{2023}).
\newblock


\bibitem[Xplore(2021)]%
        {tech21}
\bibfield{author}{\bibinfo{person}{Tech Xplore}.} \bibinfo{year}{2021}\natexlab{}.
\newblock \bibinfo{title}{AI chatbot becoming comforting companion in China for the lonely}.
\newblock
\newblock
\urldef\tempurl%
\url{https://techxplore.com/news/2021-08-ai-chatbot-comforting-china-lonely.html}
\showURL{%
\tempurl}
\newblock
\shownote{Accessed: 2024-06-02}.


\bibitem[Zamfirescu-Pereira et~al\mbox{.}(2023)]%
        {zam23}
\bibfield{author}{\bibinfo{person}{JD Zamfirescu-Pereira}, \bibinfo{person}{Richmond~Y Wong}, \bibinfo{person}{Bjoern Hartmann}, {and} \bibinfo{person}{Qian Yang}.} \bibinfo{year}{2023}\natexlab{}.
\newblock \showarticletitle{Why Johnny can’t prompt: how non-AI experts try (and fail) to design LLM prompts}. In \bibinfo{booktitle}{\emph{Proceedings of the 2023 CHI Conference on Human Factors in Computing Systems}}. \bibinfo{pages}{1--21}.
\newblock


\bibitem[Zhao et~al\mbox{.}(2024)]%
        {zhao24}
\bibfield{author}{\bibinfo{person}{Wenting Zhao}, \bibinfo{person}{Xiang Ren}, \bibinfo{person}{Jack Hessel}, \bibinfo{person}{Claire Cardie}, \bibinfo{person}{Yejin Choi}, {and} \bibinfo{person}{Yuntian Deng}.} \bibinfo{year}{2024}\natexlab{}.
\newblock \showarticletitle{WildChat: 1M ChatGPT Interaction Logs in the Wild}.
\newblock \bibinfo{journal}{\emph{arXiv preprint arXiv:2405.01470}} (\bibinfo{year}{2024}).
\newblock


\bibitem[Zheng et~al\mbox{.}(2023)]%
        {zhe23}
\bibfield{author}{\bibinfo{person}{Lianmin Zheng}, \bibinfo{person}{Wei-Lin Chiang}, \bibinfo{person}{Ying Sheng}, \bibinfo{person}{Tianle Li}, \bibinfo{person}{Siyuan Zhuang}, \bibinfo{person}{Zhanghao Wu}, \bibinfo{person}{Yonghao Zhuang}, \bibinfo{person}{Zhuohan Li}, \bibinfo{person}{Zi Lin}, \bibinfo{person}{Eric Xing}, {et~al\mbox{.}}} \bibinfo{year}{2023}\natexlab{}.
\newblock \showarticletitle{Lmsys-chat-1m: A large-scale real-world llm conversation dataset}.
\newblock \bibinfo{journal}{\emph{arXiv preprint arXiv:2309.11998}} (\bibinfo{year}{2023}).
\newblock


\bibitem[Zhou et~al\mbox{.}(2020)]%
        {zho20}
\bibfield{author}{\bibinfo{person}{Li Zhou}, \bibinfo{person}{Jianfeng Gao}, \bibinfo{person}{Di Li}, {and} \bibinfo{person}{Heung-Yeung Shum}.} \bibinfo{year}{2020}\natexlab{}.
\newblock \showarticletitle{The design and implementation of xiaoice, an empathetic social chatbot}.
\newblock \bibinfo{journal}{\emph{Computational Linguistics}} \bibinfo{volume}{46}, \bibinfo{number}{1} (\bibinfo{year}{2020}), \bibinfo{pages}{53--93}.
\newblock


\end{thebibliography}
\end{document}